\documentclass[aps, prd, reprint, longbibliography, nofootinbib,superscriptaddress, floatfix]{revtex4-1}

\usepackage{slashed}
\usepackage{verbatim}
\usepackage[T1]{fontenc}
\usepackage{mathbbol}
\usepackage[dvipsnames]{xcolor}
\usepackage{orcidlink}

\usepackage[normalem]{ulem}
\usepackage[english]{babel}
\usepackage{lipsum}
\usepackage{physics}
\usepackage{dcolumn}
\usepackage{tensor}
\usepackage{comment}
\usepackage{placeins}
\usepackage{graphicx,color,overpic,mathtools}
\usepackage{amsthm,amsmath,amssymb,mathrsfs}
\usepackage{braket,bm,bbm,setspace}
\usepackage{booktabs}
\usepackage{cancel}
\usepackage{float}
\usepackage{xargs}
\definecolor{myred}{RGB}{179, 27, 27}
\usepackage{hyperref}
\hypersetup{colorlinks=true, linkcolor=blue, citecolor=blue, urlcolor=blue}

\definecolor{cmr06}{HTML}{EB7308}

\begin{document}

\title{How to Play a Black Hole: Understanding Quasinormal Mode Excitation}

\author{Luc\'{i}a V\'{e}lez-Tartajo\orcidlink{0009-0001-3717-3576}}
\email{l.veleztartajo@uva.nl}
\affiliation{Gravitation Astroparticle Physics Amsterdam (GRAPPA),\\ University of Amsterdam, 1098 XH Amsterdam, The Netherlands}
\affiliation{Center of Gravity, Niels Bohr Institute, Blegdamsvej 17, 2100 Copenhagen, Denmark}
\author{Vitor Cardoso\orcidlink{0000-0003-0553-0433}}
\affiliation{Center of Gravity, Niels Bohr Institute, Blegdamsvej 17, 2100 Copenhagen, Denmark}
\affiliation{CENTRA, Departamento de F\'{\i}sica, Instituto Superior T\'ecnico -- IST, Universidade de Lisboa -- UL, Avenida Rovisco Pais 1, 1049-001 Lisboa, Portugal}
\author{Jaime Redondo-Yuste\orcidlink{0000-0003-3697-0319}}
\affiliation{Leinweber Forum for Theoretical Physics, Princeton University, Princeton NJ, USA}

\begin{abstract}
Understanding the excitation of black hole quasinormal modes is essential for interpreting gravitational-wave signals and inferring black hole properties from them. 
In this work, we investigate the excitation of the quasinormal modes of Schwarzschild black holes under controlled probes.
In the absence of a source, we find that broader initial data suppresses mode excitation, enhancing the relative contribution of the late-time tail, while data localized near the light ring strongly suppresses overtones relative to the fundamental mode. On the other hand, plunging sources give rise to two distinct ringdown stages, associated with scattered radiation generated during the particle's infall and with the particle crossing the light ring. 
Finally, we investigate how much of the signal is sourced from the strong-field region by switching off the source at some finite radius. The resulting waveform
is exponentially close -- in cutoff radius -- to that of a standard infall.
These results provide insight into the dependence of QNM excitation on the properties and dynamics of the perturbation, and offer a controlled setting in which to investigate the origin of black hole ringdown.

\end{abstract}
\maketitle


\section{Introduction}\label{sec:intro}
Black hole (BH) spectroscopy has emerged as one of the most active and rapidly developing areas within gravitational physics in recent years~\cite{Berti:2009kk,Berti:2025hly}. From a theoretical standpoint, the final stage of a BH coalescence --the ringdown-- can be described with remarkable accuracy using linear BH perturbation theory. In this framework, the signal is modelled as a superposition of exponentially damped sinusoids known as quasinormal modes (QNMs)\footnote{Linear perturbation theory predicts also a power-law tail and initial data effects, which have to date been mostly unaccounted for in data analysis~\cite{Berti:2009kk, Baibhav:2023clw, Berti:2025hly}.}. These modes have complex frequencies that depend solely on the BH parameters~\cite{Vishveshwara:1970zz, Chandrasekhar:1975zza, Teukolsky:1973ha, Press:1971wr, Leaver:1985ax, Leaver:1986gd, Andersson_1997, Berti:2009kk, Berti:2025hly}, and which can be computed with remarkable precision~\cite{Leaver:1985ax,Berti:2005ys,Berti:2025hly}.

Ringdown has now been directly observed in high signal-to-noise events, most notably GW250114, the loudest gravitational wave (GW) detection to date~\cite{LIGOScientific:2025wao, LIGOScientific:2025rid}. By analysing the information encoded in this phase, it is possible to infer properties of the remnant BH, such as its mass and spin, and to test General Relativity (GR) in a regime that is otherwise inaccessible. As detector sensitivities continue to improve, ringdown signals are becoming increasingly relevant for both parameter estimation and fundamental physics~\cite{Cardoso:2019rvt,Barausse:2020rsu,Gupta:2023,ET:2025xjr,Berti:2025hly}. With next-generation of GW detectors~\cite{LISA:2024hlh, ETScienceTeam_2011, Reitze:2019iox} it will be possible to probe the ringdown phase in greater detail, allowing to test the predictions of GR with unprecedented precision in the strong-field regime~\cite{Will:1994fb, Berti:2015itd,Cardoso:2019rvt,Berti:2025hly}.

Despite this progress, significantly less is understood about the mechanisms through which, and to which extent, these modes are excited in dynamical scenarios. In particular, the determination of their amplitudes and relative excitation remains a central open problem in BH spectroscopy~\cite{Leaver:1985ax, Leaver:1986gd, Sun:1988tz, Andersson:1995zk, Andersson_1997, Nollert:1998ys, Glampedakis:2001js, Berti:2006wq}, with recent work on it~\cite{Berti:2025hly, Baibhav:2023clw, DeAmicis:2025xuh, DeAmicis:2026wqd, Kholyavka:2026uam}. While the QNM frequencies are fully determined by the spacetime geometry, their amplitudes depend on the details of the perturbation. Therefore, the amplitudes can be predicted only when the source of the perturbation is known, which is not the case in realistic observational scenarios. Addressing this issue is crucial for the interpretation of observed signals, for the construction of accurate and reliable waveform models, and ultimately for robust tests of GR~\cite{Crescimbeni:2025kxi}.

This work aims to contribute to the understanding of the mechanisms governing the excitation of QNMs and its dependence on the properties of the perturbation. In particular, we address three central questions: under what conditions QNMs are excited, which features of the perturbation control their amplitudes, and to which extent the process can be regarded as localized in spacetime. 

Our purpose is to study very simple, controlled probes of excitation. Our experiments should be considered as the BH analogs of plucking a guitar string at different points and with different fingerpicks. We use narrow pulses of radiation, initialized at different locations. We find that wider initial data suppresses quasinormal mode excitation, suggesting that adiabatic perturbations enhance the late-time tail over ringdown. In Refs.~\cite{Vishveshwara:1970zz, Andersson:1995zk} it was already argued that for scalar perturbations wider than a certain critical value, the fundamental mode becomes exponentially suppressed. This has also been studied for gravitational perturbations in Ref~\cite{Kholyavka:2026uam}, in which it is shown that oscillatory initial data excite QNMs according to their Fourier content.

We also examine the dependence on the location of the initial perturbation, which has been less studied in the existing literature. We find that overtones are strongly suppressed relative to the fundamental mode when the initial perturbation is placed in the neighbourhood of the light ring: overtones are only excited away from the light ring.

We then extend this analysis by introducing a source driving the perturbations, considering a plunging particle of varying velocity. We identify two ringdown stages, whose duration depends on the particle's velocity. The first one is associated with radiation emitted by the particle at large distances and scattered off the light ring, while the second is triggered by the particle itself crossing the light ring. The first stage dominates the signal until the particle reaches the light ring, when the second one takes over.
With the purpose of understanding how much the source probes the strong-field region, we switch off the source in the neighbourhood of the light ring. We find that removing the source after it crosses the peak of the potential barrier produces a waveform that is exponentially close, as function of the cutoff radius, to that of the standard infall. This indicates that emission from the strong-field region is exponentially suppressed by the potential barrier, a result that is relevant to ongoing discussions about the origin of black hole ringdown~\cite{Oshita:2025qmn, Lu:2025vol, Ma:2026hcb, Sun:2026mto, Dyer:2026yex, Kuntz:2026xep, Cheung:2026gfd, Kankani:2026kst, Dyer:2026sox, Kubota:2026qjh}.

This paper is organised as follows. In Sec. \ref{sec:theo_framework} we introduce the BH perturbation theory framework. Sec. \ref{sec:num_framework} describes the numerical methods used to carry out the analysis. Sec. \ref{sec:results} presents the results, both for the initial value and sourced problems. Finally, in Sec. \ref{sec:discussion} we summarise our main results and their implications. We use geometric units $G=c=1$ throughout. 

\section{Framework}\label{sec:theo_framework}
We consider perturbations of a Schwarzschild geometry,
\begin{equation}
\text{d}s^2 = -f(r) \text{d}t^2 + f^{-1}(r)\text{d}r^2 + r^2 \text{d}\Omega_2^2\,,\label{eq:Schwarzschild_metric}
\end{equation}
with $f(r)=1-2M/r$. The spherical symmetry of the spacetime allows us to expand gravitational perturbations in tensor spherical harmonics~\cite{Mathews:1970, Zerilli:1970wzz} with polar and azimuthal indices $\ell, \, m$, and the linearized Einstein equations separate into odd- and even-parity sectors, which evolve independently at linear order. We restrict our analysis to the odd-parity sector. Following the standard Regge--Wheeler--Zerilli formalism \cite{Regge:1957td, Zerilli:1970wzz,Cardoso:2025npr}, the linearized Einstein equations then reduce to a single master equation,
\begin{equation}
    \left[\frac{\partial^2}{\partial x^2} - \frac{\partial^2}{\partial t^2} - V^{\rm RW}_\ell(x)\right]\psi_{\ell m}(t,x) = \mathcal{S}^{\text{ax}}_{\ell m}\,,
    \label{eq:RW_equation_time_domain}
\end{equation}
where
\begin{equation}
    V^{\rm RW}_\ell(r) = \left(1-\frac{2M}{r}\right)\left[\frac{\ell(\ell+1)}{r^2} +(1-s^2)\frac{2M}{r^3}\right]\,,
    \label{eq:RW_potential}
\end{equation}
with $s=0$ for scalar perturbations and $s=2$ for gravitational perturbations. Above, $x$ is the tortoise coordinate satisfying $\text{d}r=f(r)\text{d}x$.

Defining the Laplace transform of $\psi(t,x)$ with respect to $t$ as $\hat{\psi}(\omega,x)$, the Laplace transform of Eq.~\eqref{eq:RW_equation_time_domain} becomes~\cite{Cardoso:2025npr}
\begin{eqnarray}
&&\left[\frac{\partial^2}{\partial x^2} +\omega^2-V^{\rm RW}_{\ell}(x)\right]\hat{\psi}(\omega,x)=\mathcal{I}(\omega, x)\,,\label{eq:RW_equation_freq_domain} \\
&&\mathcal{I}(\omega, x)\equiv\hat{\mathcal{S}}^{\text{ax}}_{\ell m} + \left.i\omega\psi(t,x)\right|_{t=0} - \left.\partial_t\psi(t,x)\right|_{t=0}\,.
\end{eqnarray}
The solution to this equation can be written in terms of the frequency domain Green's function (GF) $\hat{G}(\omega, x,\bar{x})$ as
\begin{equation}
    \hat{\psi}(\omega ,x)=\int_{-\infty}^{\infty}\text{d}\bar{x}\ \hat{G}(\omega , x, \bar{x})\mathcal{I}(\omega ,\bar{x})\,.
    \label{eq:solution_greens_function_and_source}
\end{equation}
The GF is constructed from two linearly independent solutions, $\hat{\psi}_{\rm in/up}$, of the homogeneous version of Eq.~\eqref{eq:RW_equation_freq_domain}. These solutions are defined by imposing appropriate boundary conditions: $\hat{\psi}_{\rm in}$ behaves as a purely ingoing wave at the horizon, whereas $\hat{\psi}_{\rm up}$ corresponds to a purely outgoing wave at infinity, namely
\begin{equation}
    \hat{\psi}_{\rm in}(\omega, x)\sim\left\{ \begin{array}{cc}
          e^{-i\omega x}\quad  &\quad x\rightarrow -\infty\,, \\
          A_{\rm in}e^{-i\omega x} + A_{\rm out}e^{i\omega x} &\quad  x\rightarrow +\infty\,,
    \end{array}
     \right.
     \label{eq:ingoing_solution}
\end{equation}
\begin{equation}
     \hat{\psi}_{\rm up}(\omega, x)\sim\left\{ \begin{array}{cc}
          A'_{\rm in}e^{-i\omega x} + A'_{\rm out}e^{i\omega x} &\quad x\rightarrow -\infty\,, \\
          e^{i\omega x} &\quad x\rightarrow +\infty\,.
    \end{array}
    \right.
    \label{eq:outgoing_solution}
\end{equation}
In terms of the homogeneous solutions \eqref{eq:ingoing_solution} and \eqref{eq:outgoing_solution}, the GF takes the form~\cite{Maggiore:2018sht}
\begin{equation}
    \begin{split}
        \hat{G}(\omega , x, \bar{x}) =\frac{1}{W(\omega)} \Bigl[ \theta(x-\bar{x})\hat{\psi}_{\rm in}(\omega, \bar{x}) \hat{\psi}_{\rm up}(\omega, x) \\+ \theta(\bar{x}-x)\hat{\psi}_{\rm in}(\omega, x)\hat{\psi}_{\rm up}(\omega , \bar{x}) \Bigr] \,,
    \end{split}
    \label{eq:greens_function_psi}
\end{equation}
where  $\theta(x)$ is the Heaviside function, and $W(\omega)=2\omega A_{\rm in}$ is the Wronskian of $\hat{\psi}_{\rm in}$ and $\hat{\psi}_{\rm up}$. Although not explicitly stated, the ingoing and outgoing amplitudes $A_{\rm in/out}$ are functions of $\omega$. 

The QNM frequencies correspond to the poles of the GF \eqref{eq:greens_function_psi} or, equivalently, to the values of $\omega$ for which the Wronskian vanishes, $W(\omega)=0$. These frequencies are complex-valued, with the imaginary part describing the dissipation of the system. The longest lived mode is the fundamental mode and labelled by $n=0$, and the remaining modes, ordered by increasing damping rate, are dubbed overtones and labelled by increasing $n$. 

The time-domain solution $\psi(t,x)$ is obtained by performing the inverse Laplace transform of the GF solution in Eq.~\eqref{eq:solution_greens_function_and_source}. Its analytic structure gives rise to different contributions to the signal. In particular, at sufficiently late times, we can deform the contour of integration into the lower half of the complex frequency plane, yielding contributions associated with the direct propagation of the perturbation (prompt response), the poles of the GF (QNMs), and the branch-cut. The latter originates from the long-range behaviour of the Schwarzschild potential and gives rise to power-law tails~\cite{Price:1971fb, Leaver:1986gd}. Schematically, the waveform can be decomposed as $\psi(t,x) = \psi_{\rm prompt}(t,x) + \psi_{\rm Q}(t,x) + \psi_{\rm tail}(t,x)$. Once the initial perturbation or source-driven response has sufficiently decayed at the observation point, the prompt contribution becomes negligible and the signal is described by the QNM contribution together with the power-law tails. The QNM contribution can be expressed, at sufficiently late times, as a superposition of damped oscillations~\cite{Leaver:1986gd, Berti:2025hly}
\begin{equation}
    \psi_{\rm Q}(t,x_{\rm obs}) =\sum_{n=0}^N A_n e^{-i\omega_n (t-t_{\rm ref})} \,,
    \label{eq:QNM_expansion}
\end{equation}
where $x_{\rm obs}$ is the extraction radius, $t_{\rm ref}=x_0+x_{\rm obs}$ is our choice for a reference time, with $x_0$ the initial position of the perturbation, and $\omega_n\in\mathbb{C}$ is the frequency of the mode $n$. Note that $t_{\rm ref}$ is related to the propagation time of signals scattered around the light ring. The amplitude $A_n$ of each mode factors in a geometric term, the quasinormal mode excitation factors (QNEF) $B_n$, and an overlap integral $C_n$, as 
\begin{equation}
    A_n = B_n C_n\,,
    \label{eq:decomposition_A}
\end{equation}
with
\begin{equation}
    \begin{aligned}
        B_n=& \left.\frac{A_{\rm out}}{2\omega}\left(\frac{\text{d}A_{\rm in}}{\text{d}\omega}\right)^{-1}\right|_{\omega=\omega_n}  \, , \\
        C_n=&\int_{-\infty}^\infty \text{d}x'\ \frac{\mathcal{I}(\omega_n, x')\psi_{\rm in}(\omega_n, x')}{A_{\rm out}}
        \label{eq:overlap_integral} \, .
    \end{aligned}
\end{equation}
The overlap integral quantifies how efficiently the perturbation excites each QNM~\cite{Kholyavka:2026uam}, and depends on the details of the perturbation. Therefore, the amplitudes can be predicted only when the source of the perturbation is known. In realistic scenarios the initial perturbation is not known and QNM amplitudes must be directly inferred from the ringdown signal.

We can find the radial profile of the QNMs using Leaver's method~\cite{Leaver:1986gd}. By factoring out the asymptotic behaviour of the QNMs, i.e. writing
\begin{equation}
    \hat{\psi}_{\rm in}(\omega_n, x)\equiv A_{\rm out} e^{i\omega_n x} \psi^{\rm L}_n(x)\,,
\end{equation}
there exists a convergent Frobenius series for $\psi^{\rm L}_{n}$,
\begin{equation}
    \psi_n^{\rm L}(x)= \left(1-\frac{2M}{r}\right)^{-2iM\omega_n} \frac{ \sum_{j=0}^\infty a_j\left(1-\frac{2M}{r}\right)^j}{\sum_{j=0}^\infty a_j}\,,
\end{equation}
where the coefficients $a_j=a_j(\omega_n)$ satisfy
\begin{equation}
    \alpha_n a_{n+1} + \beta_n a_n + \gamma_n a_{n-1}=0\,,
\end{equation}
with
\begin{equation}
    \begin{aligned}
        \alpha_n = & n^2+(2p+2)n + (2p+1)\,, \\
        \beta_n = & -\left[2(n+2p)^2 + 2(n+2p) + \ell(\ell+1) - s^2+1\right]\,, \\
        \gamma_n = & n^2+4pn + 4p^2-s^2\,,
    \end{aligned}
\end{equation}
with $p=-2iM\omega_n$ and $s$ the spin of the perturbation. Then, $C_n$ in Eq.~\eqref{eq:overlap_integral} simplifies to
\begin{equation}
    C_n=\int_{-\infty}^\infty \text{d}x'\ \mathcal{I}(\omega_n, x')\psi^{\rm L}_n(\omega_n, x') e^{i\omega_n x'}\,.
    \label{eq:overlap_radial_function}
\end{equation}
The combination of Eqs.~\eqref{eq:decomposition_A}, \eqref{eq:overlap_integral} and \eqref{eq:overlap_radial_function} provides a frequency domain prediction of the QNM amplitudes, provided that the initial data and the source of the perturbation are known. Alternatively, the amplitudes $A_n$ can also be obtained by fitting QNMs to the time domain solutions of the RW equation. Below, we show that both approaches provide consistent results. Throughout this work, we refer to the stationary (asymptotic) QNM amplitudes. We do not consider their possible time-dependent evolution during the ringdown, which has been investigated in Refs.~\cite{DeAmicis:2025xuh, DeAmicis:2026wqd}.

\section{Methods}\label{sec:num_framework}
\subsection{Time Evolution}
We use the method of lines to solve numerically the RW equation~\eqref{eq:RW_equation_time_domain}. We evolve in time using a explicit fourth-order Runge-Kutta scheme (RK4), and a summation by parts fourth-order finite difference approximation for both the first- and second-order derivatives~\cite{Calabrese_2004}. Moreover, to avoid noise coming from high frequency modes that can grow in time at a fixed resolution, we also add Kreiss-Oliger dissipation to the algorithm~\cite{GustaffsonKreissOliger1995}. 

We impose boundary conditions that enforce a purely ingoing wave at the horizon (inner boundary of the grid) and purely outgoing at infinity (outer boundary of the grid). To prevent numerical noise from contaminating the signal, we place the boundaries far enough so that any reflected waves cannot interfere with the scattered wave in the time interval that we consider. We set $(x_{\rm min}, x_{\rm max})=(-150M, 350M)$ and evolve the initial wave packet until $t_{\rm max}=400M$. In order to resolve sufficiently small spatial features, we choose a grid spacing of $\Delta x=0.005M$ and $\Delta t=0.5\Delta x$, satisfying the Courant-Friedrichs-Lewy condition. We extract the signal far from the BH ($r_{\rm obs}=125M$).

\subsection{Inferring QNM Amplitudes}
We extract the QNM amplitudes from the time-domain signals for different initial data. Based on Eq.~\eqref{eq:QNM_expansion} we model the signal as a superposition of QNMs, and consider only the fundamental mode and two overtones, so we truncate the sum at $N=2$. We also add the power-law tail contribution. Therefore, we approximate the ringdown signal by
\begin{equation}
    \begin{split}
        y_{\rm fit}(t)= \sum_{n=0}^2 A_n\cos{[-i\omega_n(t-t_{\rm ref})]} \\+ \sum_{k=2\ell}^{2\ell +3} A_k (t-t_{\rm ref})^{-k}\,,
    \end{split}
    \label{eq:fitting_model}
\end{equation}
with $\omega_n$ the corresponding QNM frequency~\cite{CoGdata}. Note that we do not fix the power of the tail to the theoretical value $p=2\ell+3$, since this only applies asymptotically. By fitting multiple powers, we allow the algorithm to approximate a more complicated non-oscillatory contribution over the chosen fitting interval and absorb residual late-time contamination while extracting QNM amplitudes. Therefore, the parameters that we are fitting for are the amplitudes $A_i$ of both the QNMs and the tail contributions, which are complex numbers, as explained in Sec. \ref{sec:theo_framework}.

We use the matrix pencil method (MPM)~\cite{18710, 56027, Berti:2007dg} to extract the amplitudes. This method models the signal as a finite sum of exponentials, as shown in Eq. \eqref{eq:QNM_expansion}. By embedding the signal into time-shifted Hankel matrices, the method exploits the shift-invariant structure of the exponential sum to recast the QNM frequency estimation as an eigenvalue problem. We build these matrices in terms of trigonometric functions instead of exponentials, adding the corresponding terms for the tails and fit for the amplitudes. We choose the pencil parameter to be $L=3$, corresponding to the three QNMs included in our fit, namely the fundamental mode and the first two overtones.

A key property of the model is that it is linear in the amplitudes. Note that the frequencies of the QNMs are fixed by GR, and are not a free-parameter of our model, since they do not depend on the source or the initial conditions (we use the tabulated values from~\cite{CoGdata}). We can extract the amplitudes from the time-domain signal by solving an overdetermined system using the Moore-Penrose pseudoinverse, computed via singular value decomposition. This yields a least-squares estimate of the amplitudes without requiring non-linear optimisation or initial guesses, therefore making the process fast and robust. A more detailed description of the method can be found in Appendix~\ref{appendix:MPM}.

To quantify the quality of the fit, we define the mismatch between the data $y$ and the fit $y_{\text{fit}}$ as
\begin{equation}
    \mathcal{M}=1-\frac{\left<y|y_{\text{fit}}\right>}{\sqrt{\left<y|y\right>\left<y_{\rm fit}|y_{\rm fit}\right>}}\,,
    \label{eq:mismatch}
\end{equation}
where the symbol $\left<\cdot|\cdot\right>$ denotes the product defined as
\begin{equation}
    \left<f|g\right>=\int_{t_0}^{t_{\text{end}}} \text{d}t' f(t')g(t')\,.
\end{equation}

\subsection{Initial Value Problem} \label{sec:initial_value_problem}
When studying the initial value problem, we set the source in Eq.~\eqref{eq:RW_equation_time_domain} to zero. The initial value problem is illustrative to understand the response of the BH, and in particular, to parametrise the QNM amplitudes for a given family of initial data. We study two sets of initial conditions,
\begin{align}
{\bf IC1:} \quad \psi|_{t=0}=0, \ \ \partial_t\psi|_{t=0}=\frac{1}{\sigma\sqrt{2\pi}}\text{exp}\left[{-\frac{(x-x_0)^2}{2\sigma^2}}\right]\,,
    \label{eq:IC1}
\end{align}
\begin{align}
{\bf IC2:} \quad  \psi|_{t=0}=\frac{1}{\sigma\sqrt{2\pi}}\text{exp}\left[{-\frac{(x-x_0)^2}{2\sigma^2}}\right], \ \ \partial_t\psi|_{t=0}=0\,,
    \label{eq:IC2}
\end{align}
where $\sigma$ is the width of the Gaussian pulse and $x_0$ is its initial position. We perform a time-evolution and extract the time-domain signal far from the BH.

\subsection{Source Driven Problem}
We also consider the excitation produced by an infalling particle with a fixed velocity $v$. Treating the velocity as a free parameter provides a controlled way to study how the dynamics of the source itself affect the resulting waveform and the excitation of QNMs. Rather than solving the full gravitational perturbation problem -- which would have to include the stress tensor of the mechanism fixing the source velocity -- we adopt a simplified model in which a scalar field is sourced by a point-particle carrying unit scalar charge moving along a non-geodesic trajectory.

In this scenario, the scalar perturbations are ruled by Eq. \eqref{eq:RW_equation_time_domain} with $s=0$ and 
\begin{equation}
\mathcal{S}^{\rm ax}_{\ell m}=-4\pi r T, \quad T=\frac{-m_p}{\sqrt{-g}u^t}\delta^{(3)}(\mathbf{x}-\mathbf{x}_p(t))\,,
\label{eq:trace_EM_tensor_pp}
\end{equation}
where $m_p$ is a scalar charge~\cite{Misner:1972jf, Redondo-Yuste:2023snb}. Using Eq.~\eqref{eq:trace_EM_tensor_pp} and following~\cite{Cardoso:2024jme}, we model the source as
\begin{equation}
    \mathcal{S}=f(r)\frac{\delta(x-x_p(t))}{r}\,,
    \label{eq:particle_source_TM}
\end{equation}
where $x_p(t)=x_0 + vt$ is the trajectory followed by the source, with $x_0$ the initial position and $v$ is the velocity ($v<0$ for an infalling source). This expression is exact up to a rescaling of the scalar charge of the particle. For the numerical implementation, we regularize the delta-function in Eq.~\eqref{eq:particle_source_TM} with a narrow Gaussian of width $0.1M$, and place the source initially at $x_0=30M$. To isolate the contributions of the source we adopt trivial initial data. Since the particle starts at a finite radius, there is a small burst of spurious ``junk radiation'' at early times. In our simulations, this transient precedes the prompt response and has a negligible effect on the subsequent physical evolution.
 
\section{Results} \label{sec:results}

\begin{figure}
    \centering
    \includegraphics[scale=0.57]{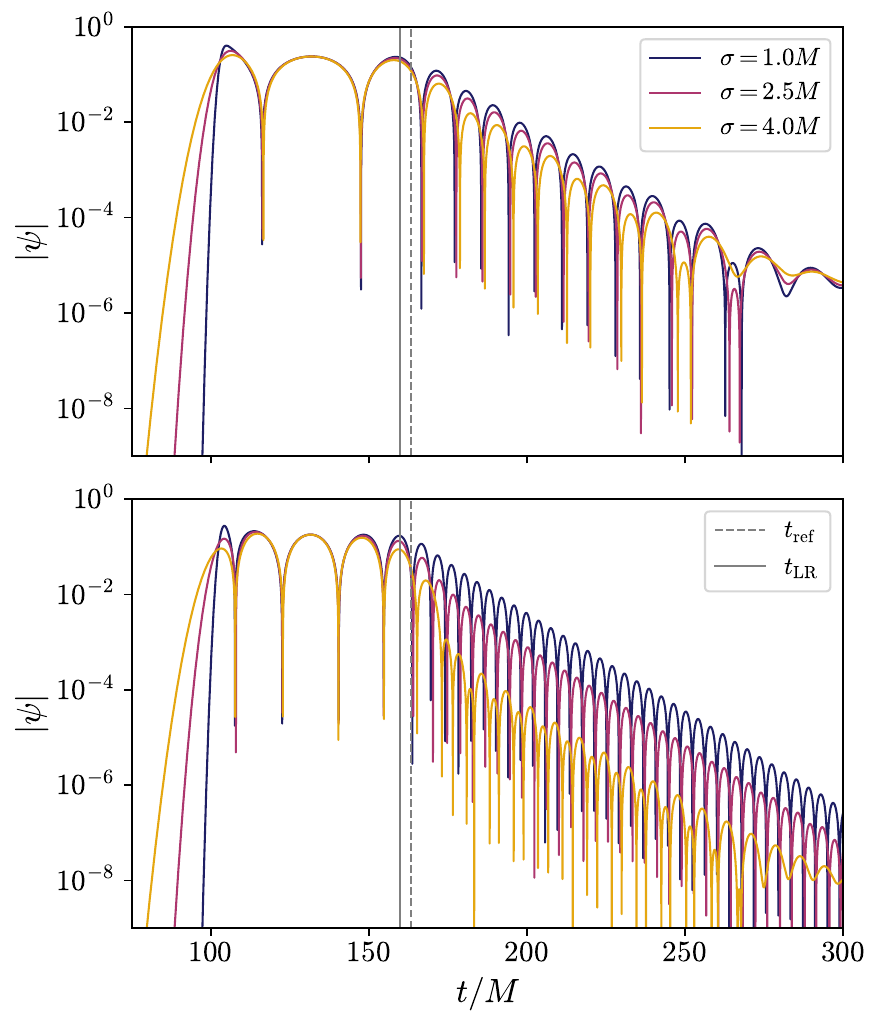}
    \caption{Ringdown signals extracted at $r_{\rm obs}=125M$ for initial conditions IC1 with $x_0=30.0$ and varying $\sigma$. On the top (bottom) we plot the signal produced by a perturbation with $\ell = 2$ ($\ell=4$). The vertical dashed line marks the reference time $t_{\rm ref}=x_0+x_{\rm obs}$ used for the fits, which is also a good reference for when the observer sees signals that interacted with the light ring. The vertical solid line marks the instant at which the IC reached the light ring.
    }
    \label{fig:waveforms_l_2_4}
\end{figure}
Fig.~\ref{fig:waveforms_l_2_4} shows some of the numerical solutions obtained using the methods presented in Sec.~\ref{sec:num_framework} for the initial value problem. They can divided into three contributions: a prompt response, exponentially damped oscillations, and a power-law tail.

The first part of the signal corresponds to the signal emitted before the perturbation reaches the light ring (left of the solid dashed line in Fig.~\ref{fig:waveforms_l_2_4}), and is known as the prompt response of the BH. Ref.~\cite{DeAmicis:2026tus} shows that it arises from the poles at $\omega=0$ in the frequency domain, being a small-frequency response. Structurally, the prompt response is expected to have $\ell$ zeros for ingoing initial data. This is precisely what we observe for signals generated using IC1 (Fig.~\ref{fig:waveforms_l_2_4}). If we instead use IC2 (see Fig.~\ref{fig:waveforms_l_2_4_IC2} in Appendix~\ref{appendix:results_IC2}), we observe an extra peak and an extra zero in the prompt response, due to the initial condition propagating directly towards the observer.

The second part corresponds to the ringdown signal, which we take to begin\footnote{The definition of the ringdown onset is not unique and remains an ongoing discussion in the literature~\cite{Bhagwat:2017tkm, Berti:2025hly, Kuntz:2026crc}} at the grey dashed line ($t_{\rm ref}$) in Fig. \ref{fig:waveforms_l_2_4}. We can see that this part is characterised by exponentially damped sinusoids and a late-time power-law tail, as discussed in Sec.~\ref{sec:theo_framework}. In Figure \ref{fig:waveforms_l_2_4}, we can see that for $\ell=4$ and larger values of $\sigma$ (yellow line) the signal's morphology is deformed with respect to the other two. In particular, we observe a beating pattern appears in the oscillations, which is caused by interference between the QNMs and the late-time tail.

\subsection{Mode content} \label{subsec:results_mode_content}
We investigate how the QNM amplitudes depend on the Gaussian parameters $\sigma$ and $x_0$ for both IC1 and IC2, comparing the time-domain (TD) amplitudes extracted from signal fits with the frequency-domain (FD) predictions obtained from the overlap integrals in Eq.~\eqref{eq:overlap_radial_function}. The results obtained for the two sets of initial conditions are broadly similar, indicating that the dependence on $\sigma$ is controlled by spectral properties rather than on whether the initial data is static or not. Therefore, for the sake of conciseness in the following we show the results for IC1. The results for IC2 can be found in Appendix~\ref{appendix:results_IC2}.

All amplitudes are referred to $t=t_{\rm ref}$ (defined below Eq.~\eqref{eq:QNM_expansion}). For each value of $\sigma$ and $x_0$, we extract the amplitudes of the fundamental mode, $A_0$, and the first overtone, $A_1$, from a fit with $N=2$. We do not show our results for the amplitude of the second overtone, since we cannot resolve it accurately enough. We fit in a window $t\in t_{\rm ref} + [t_0,80M]$, and vary the starting time $t_0 \in [0, 10M]$, to ensure that the first overtone is still present in the signal. For each set of initial data parameters, we identify a stability interval $[t_1, t_2]$ over which the amplitudes remain approximately constant, and the extracted amplitude is computed as the average over that interval. The standard deviation in this interval gives us an estimate of the amplitude uncertainty inferred within this stability interval.

\subsubsection{Dependence on the width of the pulse}
\begin{figure}
    \includegraphics[width=0.9\linewidth]{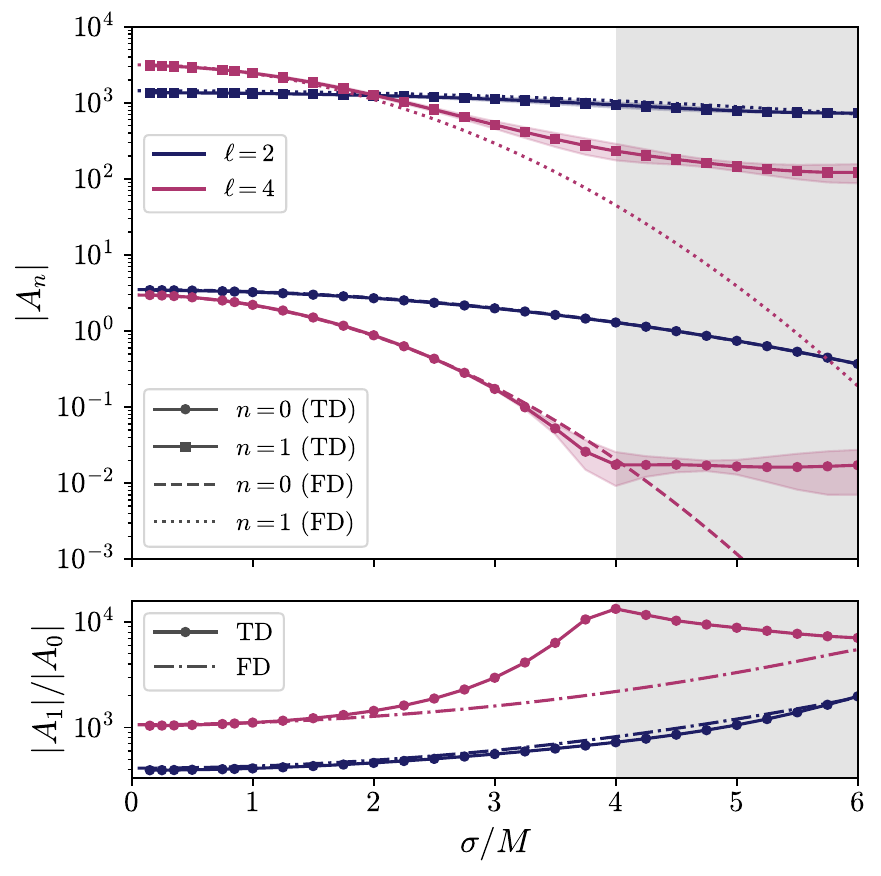}
    \caption{Amplitudes of the fundamental mode (circles) and the first overtone (squares) extracted at $r_{\rm obs}=125M$ as a function of the width $\sigma$, obtained from a fit with $N=2$ to data generated by an initial condition IC1 with $x_0=30.0M$. Blue (pink) lines correspond to data with $\ell=2$ ($\ell=4$). The data points correspond to the average values over a stable window $t_0\in[t_1, t_2],\ t_2<10M$ computed in time-domain (TD), and the coloured shaded bands correspond to the standard deviation. The dashed and dotted lines correspond to the frequency-domain (FD) calculation. The grey region ($\sigma>4M$) marks the regime for which the waveforms for $\ell=4$ become importantly deformed due to the interaction between the QNMs and the tail.
    }
    \label{fig:amplitudes_function_of_sigma}
\end{figure}
We now study how the excitation of the fundamental mode and first overtone depends on the width $\sigma$ of the initial data. Varying $\sigma$ controls the spectral bandwidth of the perturbations: narrow Gaussians probe a broad range of frequencies, whereas wider Gaussians excite a narrower band, centered at loe frequencies. This provides a clean way of studying how the QNM excitation depends on the frequency content of the initial data. 

The results are shown in Fig.~\ref{fig:amplitudes_function_of_sigma}. Over the entire range in $\sigma$ considered, the overtone amplitude exceeds that of the fundamental mode. For $\ell=2$, the amplitude of the fundamental mode decreases monotonically with increasing $\sigma$, while the overtone remains comparatively stable, leading to a steady increase in the relative excitation $A_1/A_0$. In this case, the TD amplitudes match extremely well the FD ones. For $\ell=4$, both amplitudes decrease until reaching a minimum around $\sigma\approx 4.0M$, and then increase again for larger widths, deviating from the FD results. This increase, however, is not entirely significant, since for large values of $\sigma$ the signals become very deformed, as shown in Fig. \ref{fig:waveforms_l_2_4}. This is due to the early onset of the tail and interference of higher overtones, whose amplitude is expected to increase with $\sigma$. Therefore, we should only trust the results from outside the area shaded in grey.

We conclude that the width of the initial data affects significantly the excitation of QNMs: narrow Gaussian profiles produce a higher absolute excitation of both the fundamental mode and first overtone, while wider profiles suppress the excitation. This can also be understood in terms of the spectral content of the initial data. Given that narrow Gaussians contain a broad Fourier spectrum, they tend to excite many overtones, while wide Gaussians only excite the QNMs for which the real part of the frequency overlaps with the Fourier spectrum. As a result, wide initial data can be interpreted as adiabatic perturbations, which hardly excite QNMs. As for the relative excitation, it increases with $\sigma$ due to the exponential suppression of the fundamental mode, which was predicted in~\cite{Andersson:1995zk}. Moreover, we have shown that the TD amplitudes match the FD ones.

Our results are consistent with the discussion in Ref.~\cite{Kholyavka:2026uam} about initial data $\sigma$-dependent amplitude. However, that work relies on an approximation for the initial data being placed sufficiently far away from the BH, whereas ours is generic.

\subsubsection{Dependence on the location of the pulse}
\begin{figure*}
    \includegraphics[width=\linewidth]{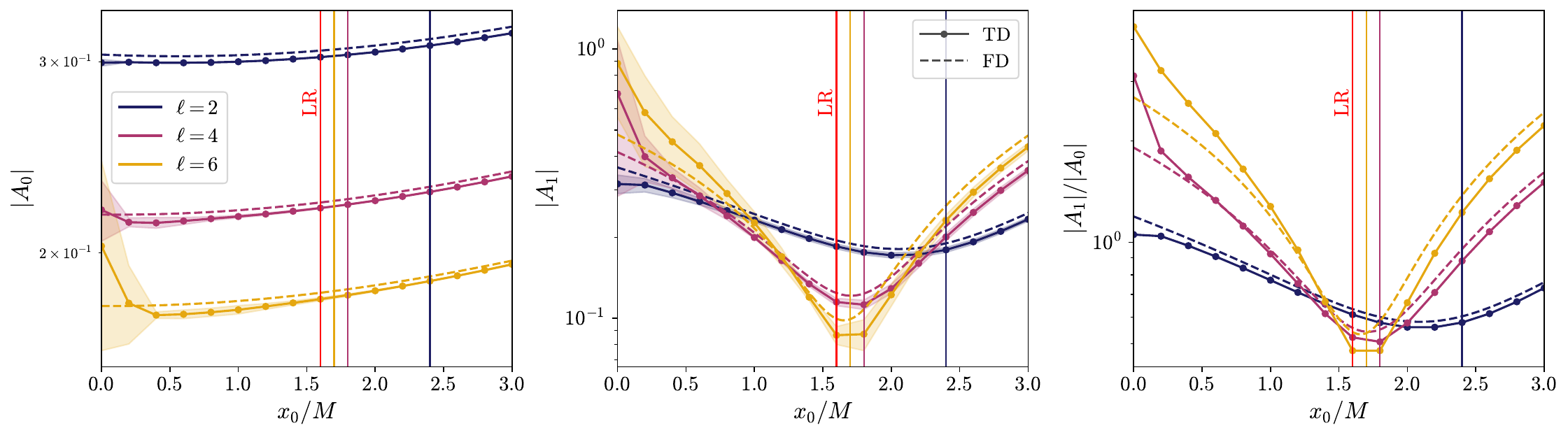}
    \caption{Amplitudes of the fundamental mode (left) and the first overtone (centre) extracted at $r_{\rm obs}=125M$ as a function of the initial position $x_0$, obtained from a fit with $N=2$ to data generated by an initial condition IC1 with $\sigma=0.05M$. The right panel shows the ratio between the first overtone and fundamental mode amplitudes. Blue, yellow, and pink lines correspond to data with $\ell=2, 4, 6$, respectively. The data points correspond to the average values over a stable window $t_0\in[t_1, t_2],\ t_2<10M$ computed in time-domain (TD), and the coloured shaded bands correspond to the standard deviation. The dashed lines correspond to the frequency-domain (FD) calculation. Vertical lines mark the position of the peak of the potential for each perturbation. The red vertical line corresponds to the location of the light ring.
    }
    \label{fig:amplitudes_function_of_x0}
\end{figure*}

While the previous subsection isolated spectral effects by varying the width $\sigma$, here we fix $\sigma$ and vary the position of the pulse in order to probe how the QNM excitation depends on where the perturbation is initially placed relative to the effective potential and light ring, in particular in the region $x_0\in[0, 3M]$. To the best of our knowledge, and despite the simplicity of this thought experiment, this has not been systematically studied before in existing literature.

To avoid artifacts associated with compact support near the boundaries and to focus on the delta-function limit, we employ narrow Gaussians with $\sigma=0.05M$. The results are displayed in Fig. \ref{fig:amplitudes_function_of_x0}. We mark with a red, solid line the location of the light ring, $x_{\rm LR}=1.6M$, and with coloured lines the peaks of the effective potential for $\ell=2, 4, 6$, at $x=2.4M, 1.8M$, and $x=1.7M$, respectively. Once more, we observe agreement between the TD and FD calculations. 

The key result is that the overtone is suppressed when the initial pulse is located in the neighbourhood of the light ring. This suppression is more precisely correlated with the peak of the corresponding effective potential than with the light ring itself. In particular, for $\ell=4, 6$ the overtone amplitude reaches a minimum coinciding with the location of the peak of the potential. For $\ell=2$ the same trend is present but less pronounced, since the overtone amplitude decreases more mildly. 

Since the peak of the effective potential approaches the light ring for increasing multipole numbers, in the eikonal limit we expect the minimum in the relative excitation to coincide with the location of the light ring. Moreover, the overtone is suppressed but not entirely absent, consistent with the finite width of the Gaussian initial data.

This suppression of the overtone can be understood in terms of the nodal structure of the QNM radial functions. In the eikonal limit, the ansatz of Ref.~\cite{Dolan:2009nk} predicts that the radial function of an overtone has a zero at the light ring, with multiplicity equal to the overtone number. Since the solution to the wave equation involves a convolution of the initial data with the QNM radial function (Sec.~\ref{sec:theo_framework}), perturbations localized near these zeros are expected to suppress the corresponding modes. This interpretation agrees with our TD and FD results. Our results also show that the peak of the potential of the perturbation has a physical impact on the excitation of overtones, unlike the bounce radius, which lies at $x=0$.

In summary, the position of the initial perturbation plays a crucial role in QNM excitation, with consistent results from the TD and FD approaches. In particular, placing the pulse near the potential peak suppresses the first overtone relative to the fundamental mode, highlighting the importance of the light ring region for QNM excitation. The TD results start to disagree with the FD near $x_0=0$. This discrepancy can be attributed to contamination from the prompt response, which is not completely removed by the fitting procedure. Indeed, as $x_0\to 0$, the reference time approaches $t_{\rm ref}=x_{\rm obs}$, making it increasingly difficult to isolate the QNM-dominated part of the signal from the prompt response. This effect is more pronounced for higher multipoles, for which the prompt response exhibits increasingly narrow temporal features.

\subsection{Source Driven Problem} \label{subsec:results_sourced_problem}
\begin{figure}
    \centering
    \includegraphics[scale=0.55]{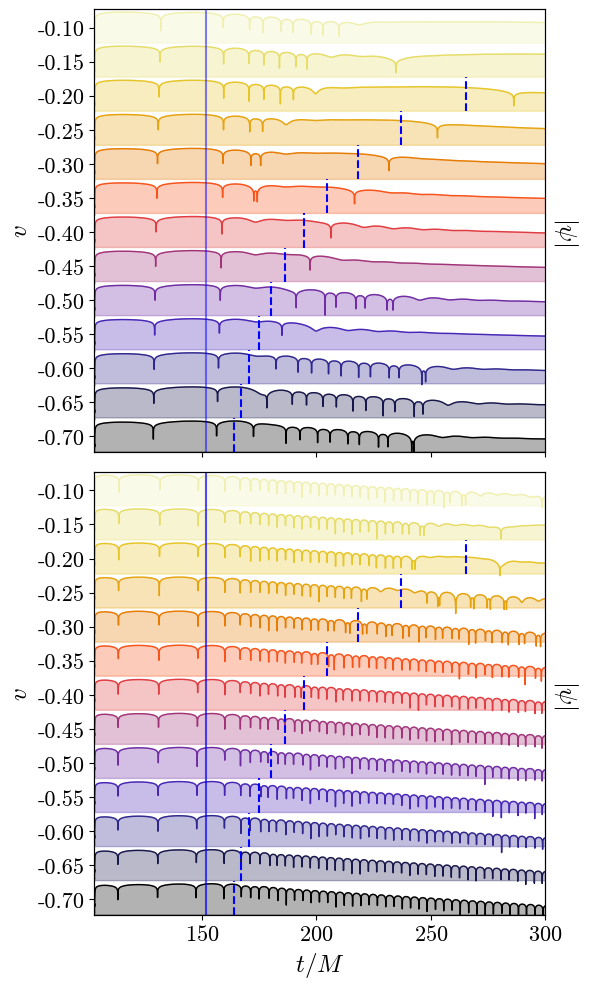}
    \caption{Numerical solutions of the sourced scalar perturbation \eqref{eq:RW_equation_time_domain} with $s=0$ for a radially infalling particle with different velocities (rows) initially placed at $x_0=30M$, extracted at $r=125M$ for a perturbation with $\ell=2$ ($\ell=4)$ on the upper (lower) panel. The vertical, dashed lines mark the instant of time in which the particle crosses the light ring, whereas the solid line marks the corresponding crossing time for a particle moving at the speed of light.
    }
    \label{fig:snapshots_l_2_v}
\end{figure}
We turn now to the source driven problem, i.e. Eq.~\eqref{eq:RW_equation_time_domain} sourced as in \eqref{eq:particle_source_TM}. Results for the scalar amplitude are summarized in Fig.~\ref{fig:snapshots_l_2_v} for different values of the velocity $v$. The top (bottom) panel shows solutions obtained for perturbations with $\ell=2$ ($\ell=4$). The vertical dashed lines mark the instant at which the corresponding particle crosses the light ring, given by
\begin{equation}
    t_{\rm cross} = \frac{x_0-x_{\rm LR}}{|v|} + x_{\rm obs} - x_{\rm LR},\,
\end{equation}
where $x_0$ is the initial position of the particle, $x_{\rm LR}$ is the location of the light ring, and $x_{\rm obs}=x(r_{\rm obs})$ with $r_{\rm obs}=125M$ is the extraction radius. 

\begin{figure}
    \centering
    \begin{overpic}[width=\linewidth]{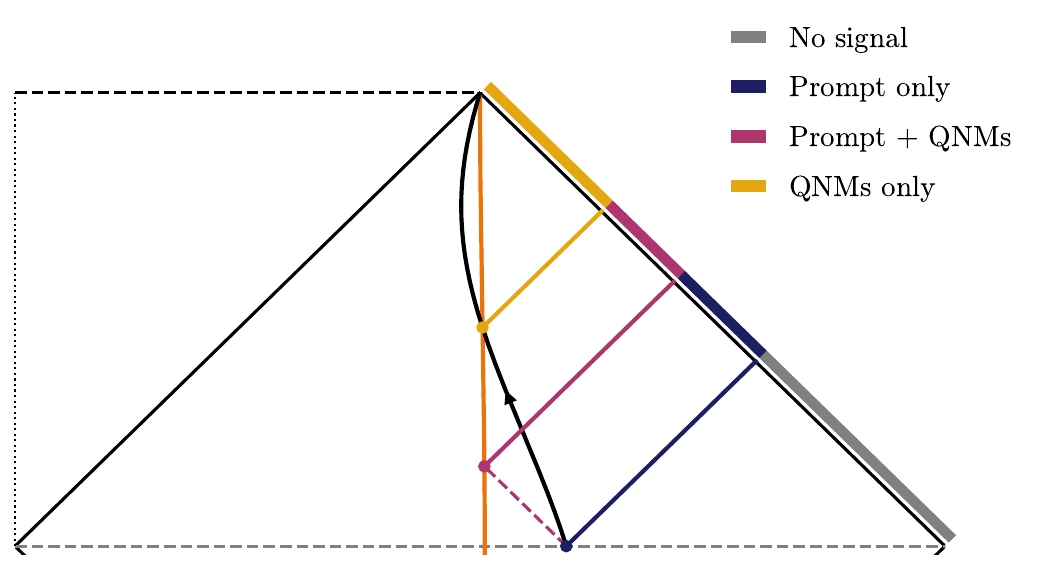}
        \put(18, 48){$r=0$}
        \put(15, 25){$\mathscr{H^+}$}
        \put(75, 25){$\mathscr{I^+}$}
        \put(45, 48){$i^+$}
        \put(93,  1){$i^0$}
        \put(57, -1){$x_0$}
        \put(38, 16){\textcolor{cmr06}{$x_{\rm LR}$}}
    \end{overpic}
    \caption{Penrose diagram of a Schwarzschild spacetime, with the worldline followed by a radially infalling particle with $v=-0.3$. The orange line marks the location of the light ring, $x_{\rm LR}$. Solid coloured lines are outgoing null rays emitted from three events: initial position (blue), light ring crossing of the radiation emitted when motion starts (pink), and light ring crossing of the particle (yellow). The coloured segments in $\mathscr{I^+}$ show the composition of the signal.
    }
    \label{fig:penrose_diag_decomposition}
\end{figure}

The signals obtained present a structure with two ringdown stages. The first stage precedes the light ring crossing of the particle, and is sourced by radiation produced as the particle travels towards the potential barrier. This radiation scatters off the potential and generates a ringdown quantitatively similar to those in Sec.~\ref{subsec:results_mode_content}. The second stage is triggered by the particle crossing the light ring itself. For high particle velocities the first stage is very brief and thus hardly observable. Fig.~\ref{fig:penrose_diag_decomposition} shows the signal observed at $\mathscr{I^+}$. The first ringdown stage corresponds to the blue and pink segments in the Figure, while the second stage corresponds to the yellow segment. The faster the particle infalls (higher $|v|$), the shorter the pink segment becomes. Note that the signal before the solid blue line is identical in all cases.

For low velocities, e.g. $|v|<0.3$, Fig.~\ref{fig:snapshots_l_2_v} shows that both ringdown stages are clearly present, but that the mode excitation in the second is suppressed for low velocities $|v|$, especially for $\ell=2$. At sufficiently low velocities the particle crosses the light ring quasi-adiabatically, so the perturbation it induces is too weak to excite QNMs appreciably. This situation is analogous to having wide initial data, which hardly excites any modes overtones.

As the velocity $|v|$ decreases, a series of low-frequency oscillations appear at the end of the first ringdown stage. These might be due to interference between the back-scattering tail of the first stage and the radiation emitted by the particle as it moves towards the light ring, and would correspond to the pink segment in Fig.~\ref{fig:penrose_diag_decomposition}. This segment becomes longer for smaller $|v|$, which allows for said interference to become more visible.

\subsubsection*{How much strong field can we probe?}
\begin{figure}
    \centering
    \includegraphics[width=\linewidth]{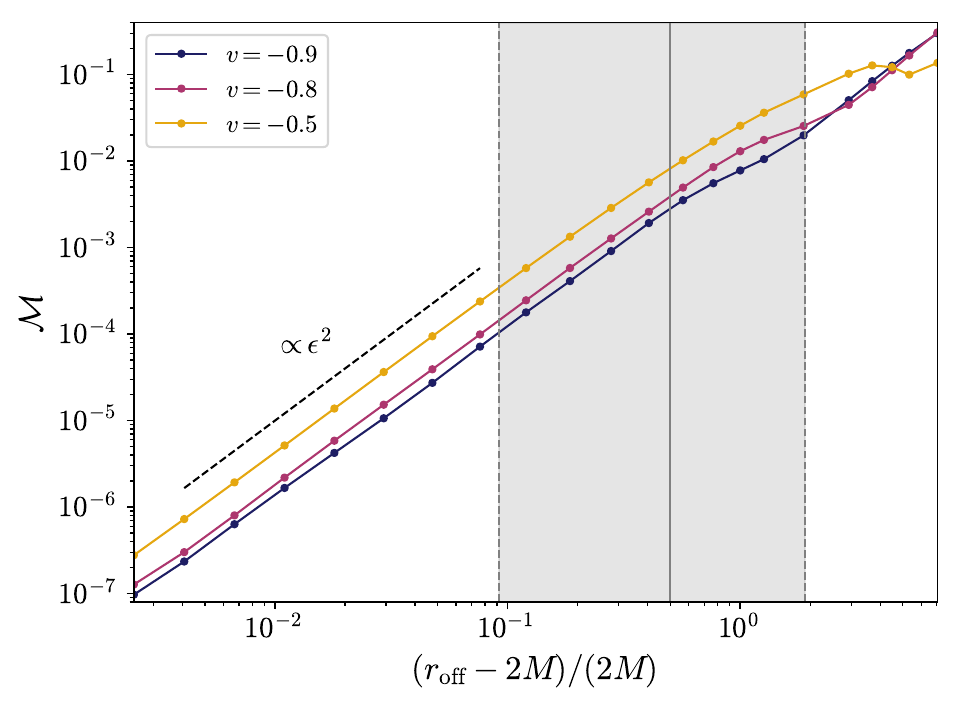}
    \caption{Mismatch as function of $\epsilon=(r_{\rm off}-2M)/(2M)$. The mismatch is calculated between the signal computed with the source present through the whole evolution and the signal in which the source is turned off at a position $r_{\rm off}$, for particles with $v\in\{-0.9, -0.8, -0.5\}$ initially placed at $x_0=30M$. The signals correspond to perturbations with $\ell=2$. The solid, grey line marks the position of the light ring, and the shaded region corresponds to the full width at half maximum of the scalar potential.
    }
    \label{fig:mismatch_vs_xoff}
\end{figure}

An important question that surfaces occasionally concerns the role of the strong-field region in radiation emission~\cite{Davis:1972ud,Cardoso:2021sip}: how much of the radiation is released when the particle is close to the horizon?

We can easily address this issue by switching off the source at a given (tortoise) coordinate position $r_{\rm off}$ ($x_{\rm off}$). This models the scenario in which the particle disappears at that position. In this case, we replace the source in Eq. \eqref{eq:particle_source_TM} by
\begin{equation}
    \mathcal{S}_{\ell m}^{\rm off}=f(r)\frac{\delta(x-x_0+vt)}{r}\frac{1}{1+e^{-a(x-x_{\rm off})}}\,,
    \label{eq:source_turnoff}
\end{equation}
where the coefficient $a=0.01M^{-1}$, so that the suppression is spatially sharp compared to the width of the potential barrier. The smoothing factor that we have added extinguishes the source for $x<x_{\rm off}$, therefore keeping the source for $x>x_{\rm off}$. For each velocity $v$ and each value of $x_{\rm off}$, we solve Eq. \eqref{eq:RW_equation_time_domain} and compute the mismatch \eqref{eq:mismatch} between this signal and the reference signal in which the source is active throughout the entire infall. This quantifies precisely how much of the GW emission is due to the source in the strong-field region. A similar problem, though specifically to the emission region of the direct wave, was considered recently in~\cite{Cheung:2026gfd}.

Fig.~\ref{fig:mismatch_vs_xoff} shows the results for perturbations with $\ell=2$ for particles with $v\in\{-0.9, -0.8, -0.5\}$ and initial position $x_0=30M$. We plot the mismatch $\mathcal{M}$ as a function of $\epsilon=(r_{\rm off}-2M)/(2M)$. When the source is switched off outside the potential barrier (right of the shaded region marking the potential barrier), the mismatch is large, of order $\mathcal{M}\gtrsim 10^{-2}$. Inside the barrier but near its peak the mismatch falls steeply, and for $\epsilon<10^{-1}$ (well inside the barrier) it decays as a power law in $\epsilon$ (or, equivalently, as an exponential in $x_{\rm off}$). In particular, we find that the behavior of the mismatch as function of truncation radius is approximately the same for all three velocities, namely $\mathcal{M}\sim\epsilon^{1.9}$ (or $\mathcal{M}\sim e^{0.95 x_{\rm off}}$). A similar result is obtained for a perturbation with $\ell=4$.

This scaling can be understood in terms of the total redshift of the particle as seen from infinity. The mismatch is quadratic in the difference between the waveforms $\Delta\psi=\psi_{\rm full} - \psi_{\rm off}$. For an observer at large $x_{\rm obs}$ and considering that $x_{\rm off}$ lies well inside the potential barrier, the difference signal becomes
\begin{equation}
    \Delta\hat{\psi}(\omega)=\frac{Ae^{i\omega x_{\rm obs}}}{vW(\omega)} e^{i\omega x_0/v} e^{-1} \int_{-\infty} ^{x_{\rm off}} \text{d}x\, e^{\left(2\kappa-i\omega \frac{1+v}{v}\right) x}\,,
\end{equation}
where $A$ is a constant, $\kappa=1/(4M)$ is the surface gravity, and $W(\omega)$ is the Wronskian defined below Eq.~\eqref{eq:greens_function_psi}. The previous equation converges at the horizon because of the factor $e^{2\kappa x}$. This is the redshift suppressing the contribution of the region close to the horizon. Integrating back to time domain and using the retarded time $u=t-x_{\rm obs}$ gives
\begin{equation}
    \Delta\psi(u)= f(x_{\rm off}) F(u-u_{\rm off})\,,
\end{equation}
with
\begin{equation}
    F(u-u_{\rm off}) = \frac{A}{2\pi} \int \text{d}\omega\frac{e^{-i\omega(u-u_{\rm off})}}{W(\omega)\left[2\kappa v -i\omega(1+v)\right]}\,,
\end{equation}
where the shape of $\Delta\psi$, namely $F(u-u_{\rm off})$ does not depend on $x_{\rm off}$ and $f(x)\to e^{2\kappa x-1}$ for $x\to -\infty$. The mismatch is quadratic in the difference between the waveforms, this suggests that for $x_{\rm off}\to -\infty$, $\mathcal{M}\propto f(x_{\rm off})^2 \propto e^{x_{\rm off}/M}$, in good agreement with the numerically observed exponent $0.95$. Changing variables to $\epsilon= (r_{\rm off}-2M)/(2M)$, the procedure above yields $\mathcal{M}\propto \epsilon^2$, which is also consistent with the numerical result we find.

Our results provide a complementary perspective on the ongoing discussion about direct waves~\cite{Oshita:2025qmn, Ma:2026hcb}, and their possible sensitivity to near-horizon dynamics~\cite{Lu:2025vol, Kuntz:2026xep, Cheung:2026gfd, Kankani:2026kst, Dyer:2026sox}. In agreement with~\cite{Cheung:2026gfd}, we find that the signal observed at infinity is predominantly sourced outside the light ring. Emission from the strong-field region is exponentially suppressed by the potential barrier at low frequencies and by gravitational redshift at high frequencies. Thus, in our setup, the imprint of near-horizon dynamics is strongly subleading, and the observed waveform is largely governed by the source dynamics around and outside the potential barrier.

\section{Discussion and conclusion} \label{sec:discussion}
Understanding the excitation of QNMs is essential for connecting BH perturbation theory with GW observations, and is one of the central open questions in BH spectroscopy. In this work, we investigate how the properties of the perturbation determine the amplitude of the excited modes and whether the excitation process can be regarded as localized in spacetime. For the initial value problem, we observe that QNM excitation is governed primarily by the spectral properties of the perturbation. The amplitude of the overtone (being lower frequency than the fundamental mode) is larger, compared to the fundamental mode, for initial perturbations with a broad radial support, in agreement with~\cite{Kholyavka:2026uam}, as well as earlier work in~\cite{Andersson:1995zk}. When considering localized initial perturbations, we find that the overtone is strongly suppressed if the initial perturbation is localized near the light ring. The minima coincide exactly with the peak of the scattering potential for each $\ell$ multipole. This can be understood from the radial structure of the QNMs. Our results highlight the importance of the spatial location of the initial perturbation and suggest that the observation of overtones requires the perturbation to extend beyond the immediate neighbourhood of the light ring.

After analysing the homogeneous problem, we turn to the source-driven case. For a simplified scalar model, we show that the signal contains two ringdown stages (Fig. \ref{fig:snapshots_l_2_v}). The first stage is sourced by the radiation produced as the particle moves towards the light ring, which gets scattered by the potential. The second stage is triggered by the particle crossing the light ring itself. The intermediate regime mixes the direct emission from the particle (Matsubara modes) with the QNMs excited from initially ingoing radiation, consistently with previous analysis based on the time-domain Green's function~\cite{Motohashi:2026mbn, Arnaudo:2025uos, Su:2026fvj, Rosato:2026moe}.

Finally, we investigate how much of the ringdown signal is sourced from within the potential barrier (Fig. \ref{fig:mismatch_vs_xoff}). We find that the mismatch decreases exponentially with the radius at which we turn off the source, $x_{\rm off}$, once the source is removed after crossing the potential barrier. These results suggest that the waveform observed at infinity is largely governed by the source dynamics around and outside the potential barrier, being the near-horizon dynamics strongly subleading. This result is in agreement with Ref.~\cite{Cheung:2026gfd} and is directly relevant to the ongoing discussion about direct waves. Even if our results come from a simplified scalar model, a similar behaviour is expected for gravitational perturbations.

Overall, our results show that QNM excitation depends not only on the background geometry, but also on the spectral content, localization, and dynamical origin of the perturbation. The simplified setups considered in this work allow us to isolate and investigate specific aspects of QNM excitation, providing useful insight into the mechanisms governing it. However, they differ substantially from the equal-mass binary mergers probed by current gravitational-wave observations, limiting the direct applicability of our findings to those systems. Extending this analysis to Kerr spacetime is a natural next step, considering that astrophysical BHs are expected to rotate.

\section*{Acknowledgements}
We are grateful to Alejandro Svyatkovskyy Kholyavka and Leart Sabani for insightful discussions and valuable comments throughout the preparation of this work.
The work of L.V.T. is supported by the European Research Council (ERC) Advanced Grant De Tenebris (grant agreement No. 101198798), and V. C. acknowledges financial support provided under the European Union’s H2020 ERC Advanced Grant “Black holes: gravitational engines of discovery” grant agreement no. Gravitas-10105258. Views and opinions expressed are however those of the authors only and do not necessarily reflect those of the European Union or the European Research Council. Neither the European Union nor the granting authority can be held responsible for them.
The Center of Gravity is a Center of Excellence funded by the Danish National Research Foundation under grant No.\@ DNRF184.
We acknowledge support by VILLUM Foundation (grant no.\@ VIL37766).
V.\,C.\@ is a Villum Investigator.  
This project has received funding from the European Union's Horizon 2020 research and innovation programme under the Marie Sk{\l}odowska-Curie grant agreement No 101007855 and No 101131233.
This work is supported by Simons Foundation International and the Simons Foundation through Simons Foundation grant SFI-MPS-BH-00012593-11.

\vspace{1cm}
\appendix

\section{Extraction of quasinormal-mode amplitudes} \label{appendix:MPM}
We provide here further details on the construction of the fitting matrix used to extract the QNM amplitudes. We work with the real part of the QNM expansion~\eqref{eq:QNM_expansion}, writing
\begin{equation}
    y_{\rm Q}(T)=
    \sum_{n=0}^{N-1} e^{-\gamma_n T}
    \left[
        \alpha_n\cos(\omega_{R,n}T)
        +\beta_n\sin(\omega_{R,n}T)
    \right]\,,
    \label{eq:real_qnm_expansion}
\end{equation}
where $T=t-t_{\rm ref}$, $\omega_n=\omega_{R,n}-i\gamma_n$, and $\alpha_n$ and $\beta_n$ are real coefficients. These are related to the complex QNM amplitudes through $A_n=\alpha_n+i\beta_n$, and thus. $|A_n|=\sqrt{\alpha_n^2+\beta_n^2}$.

Including the power-law contributions, the complete fitting model can be written as
\begin{equation}
    y_{\rm fit}(T)=y_{\rm Q}(T)
    +\sum_{p=2\ell}^{2\ell+3}a_p
    \left(\frac{x_0}{T+x_0}\right)^p\,,
    \label{eq:real_fitting_model}
\end{equation}
where $a_p$ are additional fitting coefficients, such that $A_k=a_kx_0^k$ in Eq.~\eqref{eq:QNM_expansion}. Note that we have used $t-x_{\rm obs}=T+x_0$.

For a numerical signal sampled at times $T_j$, we construct a design matrix $V$ whose columns correspond to the basis functions in Eq.~\eqref{eq:real_fitting_model}. Explicitly, its elements are
\begin{equation}
\begin{aligned}
    V_{j,2n+1}
    &=e^{-\gamma_n T_j}\cos(\omega_{R,n}T_j)\,,\\
    V_{j,2n+2}
    &=e^{-\gamma_n T_j}\sin(\omega_{R,n}T_j)\,,\\
    V_{j,2N+q}
    &=\left(\frac{x_0}{T_j+x_0}\right)^{p_q}\,,
\end{aligned}
\end{equation}
where $n=0,\ldots,N-1$ and $p_q$ denotes the $q$-th tail exponent. The fitting problem can then be expressed as
\begin{equation}
    V\mathbf{c}=\mathbf{y}\,,
\end{equation}
where $\mathbf{y}$ is the numerical signal and
\begin{equation}
    \mathbf{c}=
    (\alpha_0,\beta_0,\ldots,\alpha_{N-1},\beta_{N-1},
    a_{p_1},\ldots,a_{p_{N_{\rm tail}}})^{\rm T}
\end{equation}
contains all the unknown coefficients.

We solve this overdetermined system using the singular value decomposition $V=U\Sigma W^{\rm T}$. The Moore--Penrose pseudoinverse is constructed by retaining only singular values larger than a relative threshold $\epsilon_{\rm SVD}\sigma_{\rm max}$, where $\sigma_{\rm max}$ is the largest singular value. The coefficient vector is then obtained as
\begin{equation}
    \mathbf{c}=V^{+}\mathbf{y} =W\Sigma^+ U^{\rm T} \mathbf{y}\,.
\end{equation}
Finally, the QNM amplitudes are recovered from the corresponding pairs $(\alpha_n,\beta_n)$, independently of the fitted tail coefficients.

\section{Results for IC2} \label{appendix:results_IC2}
\begin{figure}[H]
    \centering
    \includegraphics[width=\linewidth]{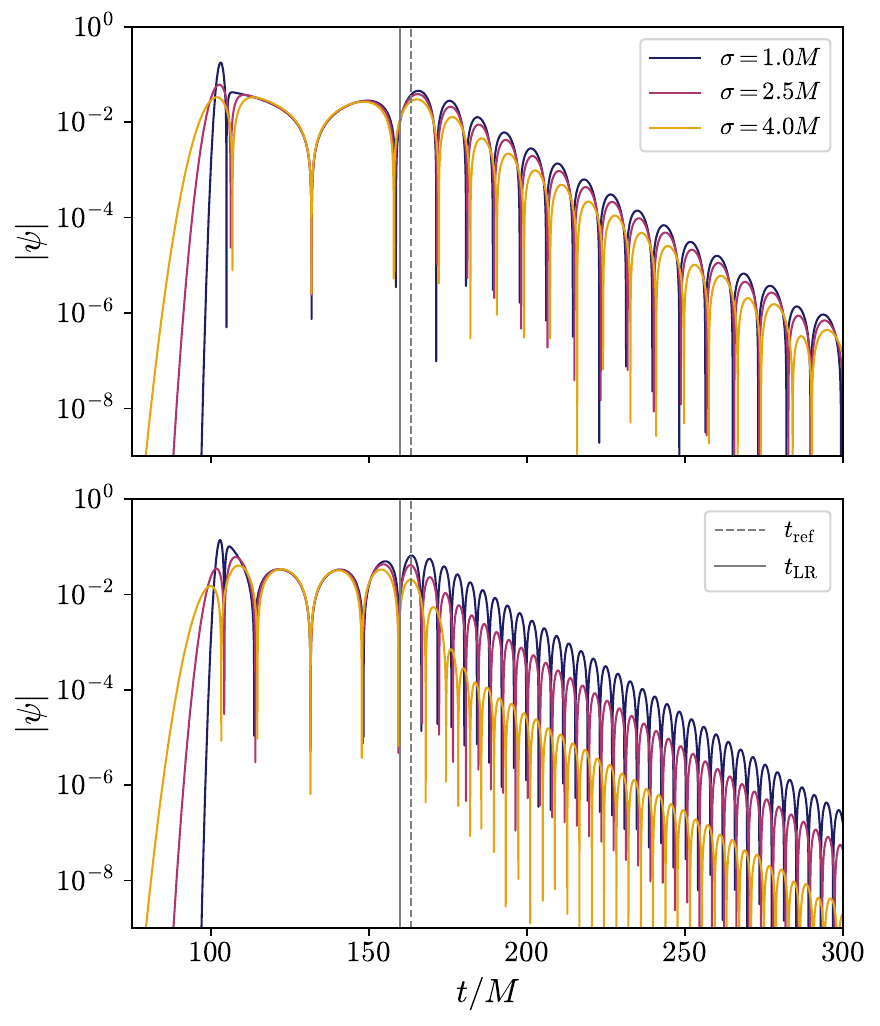}
    \caption{Ringdown signals extracted at $r_{\rm obs}=125M$ for initial conditions IC2 with $x_0=30.0$ and varying $\sigma$. On the top (bottom) we plot the signal produced by a perturbation with $\ell = 2$ ($\ell=4$). The vertical dashed line marks the reference time $t_{\rm ref}=x_0+x_{\rm obs}$ used for the fit. The vertical solid line marks the instant at which the IC reached the light ring.}
    \label{fig:waveforms_l_2_4_IC2}
\end{figure}

\begin{figure}[H]
    \centering
    \includegraphics[width=\linewidth]{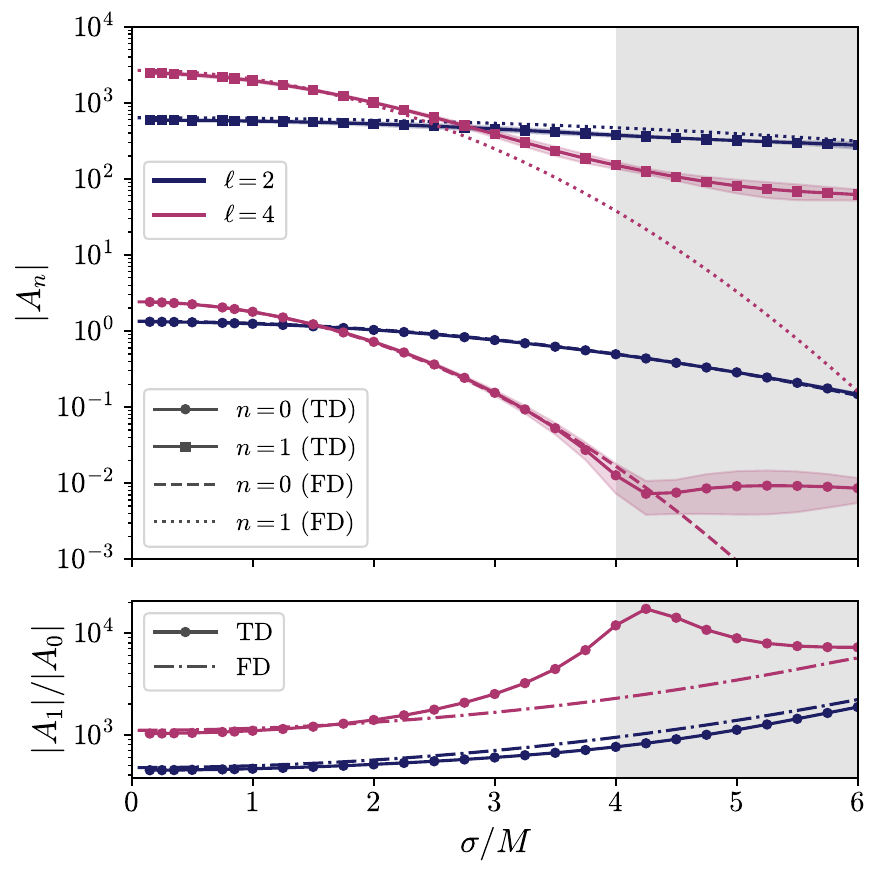}
    \caption{Amplitudes of the fundamental mode (circles) and the first overtone (squares) extracted at $r_{\rm obs}=125M$ as a function of the width $\sigma$, obtained from a fit with $N=2$ to data generated by an initial condition IC2 with $x_0=30.0M$. Blue (pink) lines correspond to data with $\ell=2$ ($\ell=4$). The data points correspond to the average values over a stable window $t_0\in[t_1, t_2],\ t_2<10M$ computed in time-domain (TD), and the shaded bands correspond to the standard deviation. The dashed and dotted lines correspond to the frequency-domain (FD) calculation. The grey region ($\sigma>4M$) marks the regime for which the waveforms for $\ell=4$ become importantly deformed due to the interaction between the QNMs and the tail.}
    \label{fig:amplitudes_function_of_sigma_IC2}
\end{figure}

\begin{figure*}
    \centering
    \includegraphics[width=\linewidth]{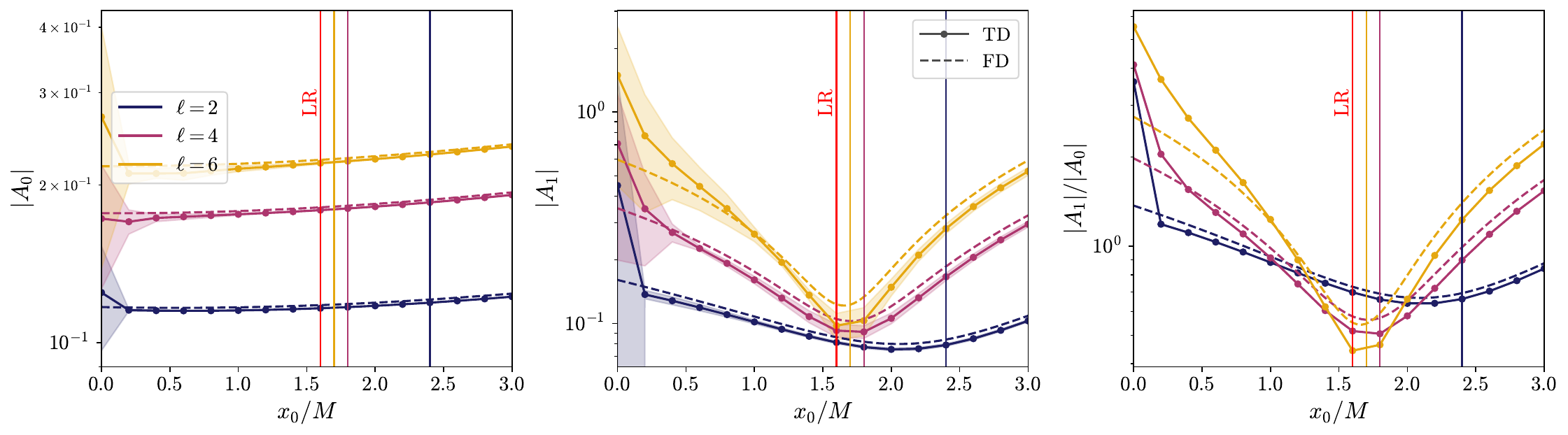}
    \caption{Amplitudes of the fundamental mode (left) and the first overtone (center) extracted at $r_{\rm obs}=125M$ as a function of the initial position $x_0$, obtained from a fit with $N=2$ to data generated by an initial condition IC2 with $\sigma=0.05M$. The right panel shows the ratio between the first overtone and fundamental mode amplitudes. Blue, yellow, and pink lines correspond to data with $\ell=2, 4, 6$, respectively. The data points correspond to the average values over a stable window $t_0\in[t_1, t_2],\ t_2<10M$ computed in time-domain (TD), and the shaded bands correspond to the standard deviation. The dashed lines correspond to the frequency-domain (FD) calculation. Vertical lines mark the position of the peak of the potential for each perturbation. The red vertical line corresponds to the location of the light ring.}
    \label{fig:amplitudes_function_of_x0_IC2}
\end{figure*}

In this section we show the results obtained for IC2. Fig.~\ref{fig:waveforms_l_2_4_IC2} shows the numerical solutions of the RW equation, Eq.~\eqref{eq:RW_equation_time_domain} for IC2. Figs.~\ref{fig:amplitudes_function_of_sigma_IC2} and~\ref{fig:amplitudes_function_of_x0_IC2} correspond to the study on the dependence of the amplitudes on the width and position of the initial data. As discussed in Sec.~\ref{subsec:results_mode_content}, the results are similar to the ones obtained for IC1.

\bibliography{biblio}

@book{Maggiore:2018sht,
  author    = "Maggiore, Michele",
  title     = "{Gravitational Waves. Vol. 2: Astrophysics and Cosmology}",
  publisher = "Oxford University Press",
  year      = "2018"
}

@article{Cardoso:2021sip,
    author = "Cardoso, Vitor and Duque, Francisco and Foschi, Arianna",
    title = "{Light ring and the appearance of matter accreted by black holes}",
    eprint = "2102.07784",
    archivePrefix = "arXiv",
    primaryClass = "gr-qc",
    doi = "10.1103/PhysRevD.103.104044",
    journal = "Phys. Rev. D",
    volume = "103",
    number = "10",
    pages = "104044",
    year = "2021"
}

@article{Davis:1972ud,
    author = "Davis, M. and Ruffini, R. and Tiomno, J.",
    title = "{Pulses of gravitational radiation of a particle falling radially into a schwarzschild black hole}",
    doi = "10.1103/PhysRevD.5.2932",
    journal = "Phys. Rev. D",
    volume = "5",
    pages = "2932--2935",
    year = "1972"
}

@article{Regge:1957td,
  author  = "Regge, Tullio and Wheeler, John A.",
  title   = "{Stability of a Schwarzschild Singularity}",
  journal = "Phys. Rev.",
  volume  = "108",
  pages   = "1063--1069",
  year    = "1957",
  doi     = "10.1103/PhysRev.108.1063"
}

@article{Baibhav:2023clw,
    author = "Baibhav, Vishal and Cheung, Mark Ho-Yeuk and Berti, Emanuele and Cardoso, Vitor and Carullo, Gregorio and Cotesta, Roberto and Del Pozzo, Walter and Duque, Francisco",
    title = "{Agnostic black hole spectroscopy: Quasinormal mode content of numerical relativity waveforms and limits of validity of linear perturbation theory}",
    eprint = "2302.03050",
    archivePrefix = "arXiv",
    primaryClass = "gr-qc",
    doi = "10.1103/PhysRevD.108.104020",
    journal = "Phys. Rev. D",
    volume = "108",
    number = "10",
    pages = "104020",
    year = "2023"
}

@article{Berti:2005ys,
    author = "Berti, Emanuele and Cardoso, Vitor and Will, Clifford M.",
    title = "{On gravitational-wave spectroscopy of massive black holes with the space interferometer LISA}",
    eprint = "gr-qc/0512160",
    archivePrefix = "arXiv",
    doi = "10.1103/PhysRevD.73.064030",
    journal = "Phys. Rev. D",
    volume = "73",
    pages = "064030",
    year = "2006"
}

@article{Cardoso:2019rvt,
    author = "Cardoso, Vitor and Pani, Paolo",
    title = "{Testing the nature of dark compact objects: a status report}",
    eprint = "1904.05363",
    archivePrefix = "arXiv",
    primaryClass = "gr-qc",
    doi = "10.1007/s41114-019-0020-4",
    journal = "Living Rev. Rel.",
    volume = "22",
    number = "1",
    pages = "4",
    year = "2019"
}

@article{Misner:1972jf,
    author = "Misner, Charles W. and Breuer, R. A. and Brill, D. R. and Chrzanowski, P. L. and Hughes, H. G. and Pereira, C. M.",
    title = "{Gravitational synchrotron radiation in the schwarzschild geometry}",
    doi = "10.1103/PhysRevLett.28.998",
    journal = "Phys. Rev. Lett.",
    volume = "28",
    pages = "998--1001",
    year = "1972"
}

@inproceedings{Cardoso:2025npr,
  author        = "Cardoso, Vitor and Biswas, Shauvik and Sarkar, Subhodeep",
  title         = "{The Physics of Black Holes and Their Environments: Consequences for Gravitational Wave Science}",
  eprint        = "2511.14841",
  archivePrefix = "arXiv",
  primaryClass  = "gr-qc",
  month         = "11",
  year          = "2025"
}

@article{Mathews:1970,
  author  = "Mathews, Jon",
  title   = "{Gravitational Multipole Radiation}",
  journal = "J. Soc. Indust. Appl. Math.",
  volume  = "10",
  number  = "4",
  pages   = "768--780",
  year    = "1962",
  doi     = "10.1137/0110059"
}

@misc{CoGdata,
  author = "{The Center of Gravity}",
  title  = "{Data and Routines}",
  url    = "https://the-center-of-gravity.com/data-and-routines/"
}

@article{18710,
  author  = "Hua, Y. and Sarkar, T. K.",
  title   = "{Generalized Pencil-of-Function Method for Extracting Poles of an EM System from Its Transient Response}",
  journal = "IEEE Trans. Antennas Propag.",
  volume  = "37",
  number  = "2",
  pages   = "229--234",
  year    = "1989",
  doi     = "10.1109/8.18710"
}

@article{Zerilli:1970wzz,
  author  = "Zerilli, F. J.",
  title   = "{Gravitational Field of a Particle Falling in a Schwarzschild Geometry Analyzed in Tensor Harmonics}",
  journal = "Phys. Rev. D",
  volume  = "2",
  pages   = "2141--2160",
  year    = "1970",
  doi     = "10.1103/PhysRevD.2.2141"
}

@article{Calabrese_2004,
  author  = "Calabrese, Gioel and Lehner, Luis and Reula, Oscar and Sarbach, Olivier and Tiglio, Manuel",
  title   = "{Summation by Parts and Dissipation for Domains with Excised Regions}",
  journal = "Class. Quant. Grav.",
  volume  = "21",
  number  = "24",
  pages   = "5735--5757",
  year    = "2004",
  doi     = "10.1088/0264-9381/21/24/004"
}

@article{Andersson:1995zk,
  author  = "Andersson, N.",
  title   = "{Excitation of Schwarzschild Black Hole Quasinormal Modes}",
  journal = "Phys. Rev. D",
  volume  = "51",
  pages   = "353--363",
  year    = "1995",
  doi     = "10.1103/PhysRevD.51.353"
}

@article{Vishveshwara:1970zz,
  author  = "Vishveshwara, C. V.",
  title   = "{Scattering of Gravitational Radiation by a Schwarzschild Black Hole}",
  journal = "Nature",
  volume  = "227",
  pages   = "936--938",
  year    = "1970",
  doi     = "10.1038/227936a0"
}

@article{Sun:1988tz,
  author  = "Sun, Y. and Price, R. H.",
  title   = "{Excitation of Quasinormal Ringing of a Schwarzschild Black Hole}",
  journal = "Phys. Rev. D",
  volume  = "38",
  pages   = "1040--1052",
  year    = "1988",
  doi     = "10.1103/PhysRevD.38.1040"
}

@article{Andersson_1997,
  author  = "Andersson, Nils",
  title   = "{Evolving Test Fields in a Black-Hole Geometry}",
  journal = "Phys. Rev. D",
  volume  = "55",
  number  = "2",
  pages   = "468--479",
  year    = "1997",
  doi     = "10.1103/PhysRevD.55.468"
}

@article{Leaver:1986gd,
  author  = "Leaver, Edward W.",
  title   = "{Spectral Decomposition of the Perturbation Response of the Schwarzschild Geometry}",
  journal = "Phys. Rev. D",
  volume  = "34",
  pages   = "384--408",
  year    = "1986",
  doi     = "10.1103/PhysRevD.34.384"
}

@article{Berti:2025hly,
  author        = "Berti, Emanuele and others",
  title         = "{Black Hole Spectroscopy: From Theory to Experiment}",
  year          = "2025",
  eprint        = "2505.23895",
  archivePrefix = "arXiv",
  primaryClass  = "gr-qc"
}

@article{Berti:2006wq,
  author        = "Berti, Emanuele and Cardoso, Vitor",
  title         = "{Quasinormal Ringing of Kerr Black Holes. I. The Excitation Factors}",
  journal       = "Phys. Rev. D",
  volume        = "74",
  pages         = "104020",
  year          = "2006",
  doi           = "10.1103/PhysRevD.74.104020",
  archivePrefix = "arXiv",
  eprint        = "gr-qc/0605118"
}

@article{Cardoso:2024jme,
  author        = "Cardoso, Vitor and Carullo, Gregorio and De Amicis, Marina and Duque, Francisco and Katagiri, Takuya and Pereniguez, David and Redondo-Yuste, Jaime and Spieksma, Thomas F. M. and Zhong, Zhen",
  title         = "{Hushing Black Holes: Tails in Dynamical Spacetimes}",
  journal       = "Phys. Rev. D",
  volume        = "109",
  number        = "12",
  pages         = "L121502",
  year          = "2024",
  doi           = "10.1103/PhysRevD.109.L121502",
  archivePrefix = "arXiv",
  eprint        = "2405.12290",
  primaryClass  = "gr-qc"
}

@article{Berti:2007dg,
  author        = "Berti, Emanuele and Cardoso, Vitor and Gonzalez, Jose A. and Sperhake, Ulrich",
  title         = "{Mining Information from Binary Black Hole Mergers: A Comparison of Estimation Methods for Complex Exponentials in Noise}",
  journal       = "Phys. Rev. D",
  volume        = "75",
  pages         = "124017",
  year          = "2007",
  doi           = "10.1103/PhysRevD.75.124017",
  archivePrefix = "arXiv",
  eprint        = "gr-qc/0701086"
}

@article{56027,
  author  = "Hua, Y. and Sarkar, T. K.",
  title   = "{Matrix Pencil Method for Estimating Parameters of Exponentially Damped/Undamped Sinusoids in Noise}",
  journal = "IEEE Trans. Acoust. Speech Signal Process.",
  volume  = "38",
  number  = "5",
  pages   = "814--824",
  year    = "1990",
  doi     = "10.1109/29.56027"
}

@article{DeAmicis:2026tus,
  author        = "De Amicis, Marina and Cannizzaro, Enrico",
  title         = "{Post-Minkowskian Expansion of the Prompt Response in a Schwarzschild Background}",
  year          = "2026",
  eprint        = "2601.11706",
  archivePrefix = "arXiv",
  primaryClass  = "gr-qc"
}

@article{Press:1971wr,
  author  = "Press, William H.",
  title   = "{Long Wave Trains of Gravitational Waves from a Vibrating Black Hole}",
  journal = "Astrophys. J. Lett.",
  volume  = "170",
  pages   = "L105--L108",
  year    = "1971",
  doi     = "10.1086/180849"
}

@book{GustaffsonKreissOliger1995,
  author    = "Gustaffson, B. and Kreiss, H. and Oliger, J.",
  title     = "{Time Dependent Problems and Difference Methods}",
  publisher = "Wiley",
  address   = "New York",
  year      = "1995"
}

@article{Dolan:2009nk,
  author        = "Dolan, Sam R. and Ottewill, Adrian C.",
  editor        = "Uranga, A. M.",
  title         = "{On an Expansion Method for Black Hole Quasinormal Modes and Regge Poles}",
  journal       = "Class. Quant. Grav.",
  volume        = "26",
  pages         = "225003",
  year          = "2009",
  doi           = "10.1088/0264-9381/26/22/225003",
  eprint        = "0908.0329",
  archivePrefix = "arXiv",
  primaryClass  = "gr-qc"
}

@article{Redondo-Yuste:2023snb,
  author        = "Redondo-Yuste, Jaime and Cardoso, Vitor and Macedo, Caio F. B. and van de Meent, Maarten",
  title         = "{Eternal Binaries}",
  journal       = "Phys. Rev. D",
  volume        = "107",
  number        = "12",
  pages         = "124025",
  year          = "2023",
  doi           = "10.1103/PhysRevD.107.124025",
  eprint        = "2304.02039",
  archivePrefix = "arXiv",
  primaryClass  = "gr-qc"
}

@article{LISA:2024hlh,
  author        = "Colpi, Monica and others",
  collaboration = "LISA",
  title         = "{LISA Definition Study Report}",
  eprint        = "2402.07571",
  archivePrefix = "arXiv",
  primaryClass  = "astro-ph.CO",
  month         = "2",
  year          = "2024"
}

@book{ETScienceTeam_2011,
  author       = "{ET Science Team}",
  title        = "{Einstein Gravitational Wave Telescope Conceptual Design Study}",
  institution  = "Zenodo",
  year         = "2011",
  doi          = "10.5281/zenodo.3911260"
}

@article{Reitze:2019iox,
  author        = "Reitze, David and others",
  title         = "{Cosmic Explorer: The U.S. Contribution to Gravitational-Wave Astronomy beyond LIGO}",
  eprint        = "1907.04833",
  archivePrefix = "arXiv",
  primaryClass  = "astro-ph.IM",
  reportNumber  = "LIGO-P1900316",
  journal       = "Bull. Am. Astron. Soc.",
  volume        = "51",
  number        = "7",
  pages         = "035",
  year          = "2019"
}

@article{LIGOScientific:2025wao,
  author        = "Abac, A. G. and others",
  collaboration = "LIGO Scientific, Virgo, KAGRA",
  title         = "{Black Hole Spectroscopy and Tests of General Relativity with GW250114}",
  eprint        = "2509.08099",
  archivePrefix = "arXiv",
  primaryClass  = "gr-qc",
  reportNumber  = "LIGO P2500461",
  doi           = "10.1103/6c61-fm1n",
  journal       = "Phys. Rev. Lett.",
  volume        = "136",
  number        = "4",
  pages         = "041403",
  year          = "2026"
}

@article{LIGOScientific:2025rid,
  author        = "Abac, A. G. and others",
  collaboration = "LIGO Scientific, Virgo, KAGRA",
  title         = "{GW250114: Testing Hawking's Area Law and the Kerr Nature of Black Holes}",
  eprint        = "2509.08054",
  archivePrefix = "arXiv",
  primaryClass  = "gr-qc",
  reportNumber  = "LIGO-P2500421",
  doi           = "10.1103/kw5g-d732",
  journal       = "Phys. Rev. Lett.",
  volume        = "135",
  number        = "11",
  pages         = "111403",
  year          = "2025"
}

@article{Berti:2009kk,
  author        = "Berti, Emanuele and Cardoso, Vitor and Starinets, Andrei O.",
  title         = "{Quasinormal Modes of Black Holes and Black Branes}",
  eprint        = "0905.2975",
  archivePrefix = "arXiv",
  primaryClass  = "gr-qc",
  doi           = "10.1088/0264-9381/26/16/163001",
  journal       = "Class. Quant. Grav.",
  volume        = "26",
  pages         = "163001",
  year          = "2009"
}

@article{Will:1994fb,
  author        = "Will, Clifford M.",
  title         = "{Testing Scalar-Tensor Gravity with Gravitational Wave Observations of Inspiraling Compact Binaries}",
  eprint        = "gr-qc/9406022",
  archivePrefix = "arXiv",
  reportNumber  = "WUGRAV-94-6",
  doi           = "10.1103/PhysRevD.50.6058",
  journal       = "Phys. Rev. D",
  volume        = "50",
  pages         = "6058--6067",
  year          = "1994"
}

@article{Berti:2015itd,
  author        = "Berti, Emanuele and others",
  title         = "{Testing General Relativity with Present and Future Astrophysical Observations}",
  eprint        = "1501.07274",
  archivePrefix = "arXiv",
  primaryClass  = "gr-qc",
  doi           = "10.1088/0264-9381/32/24/243001",
  journal       = "Class. Quant. Grav.",
  volume        = "32",
  pages         = "243001",
  year          = "2015"
}

@article{ET:2025xjr,
  author        = "Abac, Adrian and others",
  collaboration = "ET",
  title         = "{The Science of the Einstein Telescope}",
  eprint        = "2503.12263",
  archivePrefix = "arXiv",
  primaryClass  = "gr-qc",
  reportNumber  = "ET-0036C-25",
  doi           = "10.1088/1475-7516/2026/03/081",
  journal       = "JCAP",
  volume        = "03",
  pages         = "081",
  year          = "2026"
}

@article{Gupta:2023,
  author        = {Gupta, Ish and Afle, Chaitanya and Arun, K. G. and Bandopadhyay, Ananya and Baryakhtar, Masha and Biscoveanu, Sylvia and Borhanian, Ssohrab and Broekgaarden, Floor and Corsi, Alessandra and Dhani, Arnab and Evans, Matthew and Hall, Evan D. and Hannuksela, Otto A. and Kacanja, Keisi and Kashyap, Rahul and Khadkikar, Sanika and Kuns, Kevin and Li, Tjonnie G. F. and Miller, Andrew L. and Nitz, Alexander Harvey and Owen, Benjamin J. and Palomba, Cristiano and Pearce, Anthony and Phurailatpam, Hemantakumar and Rajbhandari, Binod and Read, Jocelyn and Romano, Joseph D. and Sathyaprakash, B. S. and Shoemaker, David H. and Singh, Divya and Vitale, Salvatore and Barsotti, Lisa and Berti, Emanuele and Cahillane, Craig and Chen, Hsin-Yu and Fritschel, Peter and Haster, Carl-Johan and Landry, Philippe and Lovelace, Geoffrey and McClelland, David and Slagmolen, Bram J. J. and Smith, Joshua and Soares-Santos, Marcelle and Sun, Ling and Tanner, David and Yamamoto, Hiro and Zucker, Michael},
  title         = "{Characterizing Gravitational Wave Detector Networks: From A$^\sharp$ to Cosmic Explorer}",
  eprint        = "2307.10421",
  archivePrefix = "arXiv",
  primaryClass  = "gr-qc",
  year          = "2024"
}

@article{Barausse:2020rsu,
  author        = "Barausse, Enrico and others",
  title         = "{Prospects for Fundamental Physics with LISA}",
  journal       = "Gen. Rel. Grav.",
  volume        = "52",
  number        = "8",
  pages         = "81",
  year          = "2020",
  doi           = "10.1007/s10714-020-02691-1",
  eprint        = "2001.09793",
  archivePrefix = "arXiv",
  primaryClass  = "gr-qc"
}

@article{DeAmicis:2025xuh,
  author        = "De Amicis, Marina and Cannizzaro, Enrico and Carullo, Gregorio and Sberna, Laura",
  title         = "{Dynamical Quasinormal Mode Excitation}",
  journal       = "Phys. Rev. D",
  volume        = "113",
  number        = "2",
  pages         = "024048",
  year          = "2026",
  doi           = "10.1103/bgjv-lvyd",
  eprint        = "2506.21668",
  archivePrefix = "arXiv",
  primaryClass  = "gr-qc"
}

@article{DeAmicis:2026wqd,
    author = "De Amicis, Marina and Cannizzaro, Enrico and Carullo, Gregorio and Kuntz, Adrien and Sberna, Laura",
    title = "{Dynamical quasinormal mode excitation II: propagation and convergence in Schwarzschild}",
    eprint = "2605.16492",
    archivePrefix = "arXiv",
    primaryClass = "gr-qc",
    month = "5",
    year = "2026"
}

@article{Chandrasekhar:1975zza,
  author  = "Chandrasekhar, S. and Detweiler, S. L.",
  title   = "{The Quasi-Normal Modes of the Schwarzschild Black Hole}",
  journal = "Proc. Roy. Soc. Lond. A",
  volume  = "344",
  pages   = "441--452",
  year    = "1975",
  doi     = "10.1098/rspa.1975.0112"
}

@article{Leaver:1985ax,
  author  = "Leaver, E. W.",
  title   = "{An Analytic Representation for the Quasi-Normal Modes of Kerr Black Holes}",
  journal = "Proc. Roy. Soc. Lond. A",
  volume  = "402",
  pages   = "285--298",
  year    = "1985",
  doi     = "10.1098/rspa.1985.0119"
}

@article{Crescimbeni:2025kxi,
  author        = "Crescimbeni, Francesco and Forteza, Xisco Jimenez and Pani, Paolo",
  title         = "{Testing Gravity with Black Hole Ringdown Amplitudes}",
  journal       = "Phys. Rev. D",
  volume        = "113",
  number        = "4",
  pages         = "044064",
  year          = "2026",
  doi           = "10.1103/ptg5-f769",
  eprint        = "2510.11782",
  archivePrefix = "arXiv",
  primaryClass  = "gr-qc"
}

@article{Price:1971fb,
  author  = "Price, Richard H.",
  title   = "{Nonspherical Perturbations of Relativistic Gravitational Collapse. I. Scalar and Gravitational Perturbations}",
  journal = "Phys. Rev. D",
  volume  = "5",
  pages   = "2419--2438",
  year    = "1972",
  doi     = "10.1103/PhysRevD.5.2419"
}

@article{Teukolsky:1973ha,
  author  = "Teukolsky, Saul A.",
  title   = "{Perturbations of a Rotating Black Hole. I. Fundamental Equations for Gravitational, Electromagnetic, and Neutrino-Field Perturbations}",
  journal = "Astrophys. J.",
  volume  = "185",
  pages   = "635--647",
  year    = "1973",
  doi     = "10.1086/152444"
}

@article{Kholyavka:2026uam,
    author = "Kholyavka, Alejandro Svyatkovskyy and Vega, Jose Antonio Le{\'o}n and G{\'o}mez, Samuel G{\'o}mez and Forteza, Xisco Jim{\'e}nez and Datta, Sayak",
    title = "{Shaping black hole resonances: Black hole ringdown as a spectral filtering process}",
    eprint = "2605.24704",
    archivePrefix = "arXiv",
    primaryClass = "gr-qc",
    doi = "10.1103/gptl-zzcn",
    journal = "Phys. Rev. D",
    volume = "114",
    number = "6",
    pages = "064078",
    year = "2026"
}

@article{Cheung:2026gfd,
    author = "Cheung, Mark Ho-Yeuk",
    title = "{How much of the black hole ringdown is sourced within the light ring?}",
    eprint = "2608.29466",
    archivePrefix = "arXiv",
    primaryClass = "gr-qc",
    month = "8",
    year = "2026"
}

@article{Oshita:2025qmn,
    author = "Oshita, Naritaka and Ma, Sizheng and Chen, Yanbei and Yang, Huan",
    title = "{Probing Direct Waves in Black Hole Ringdowns}",
    eprint = "2509.09165",
    archivePrefix = "arXiv",
    primaryClass = "gr-qc",
    reportNumber = "YITP-25-141, RIKEN-iTHEMS-Report-25",
    month = "9",
    year = "2025"
}

@article{Lu:2025vol,
    author = "Lu, Neil and Ma, Sizheng and Piccinni, Ornella J. and Chen, Yanbei and Sun, Ling",
    title = "{GW250114 reveals signatures of post-merger black-hole horizon}",
    eprint = "2510.01001",
    archivePrefix = "arXiv",
    primaryClass = "gr-qc",
    doi = "10.1038/s41586-026-10696-0",
    journal = "Nature",
    volume = "655",
    number = "8121",
    pages = "71--75",
    year = "2026"
}

@article{Ma:2026hcb,
    author = "Ma, Sizheng and Wang, Hai-Yang",
    title = "{Foundations of Direct Waves in Schwarzschild Ringdown}",
    eprint = "2607.06677",
    archivePrefix = "arXiv",
    primaryClass = "gr-qc",
    month = "7",
    year = "2026"
}

@article{Kuntz:2026xep,
    author = "Kuntz, Adrien and Della Rocca, Matteo",
    title = "{Vanishing of all redshift modes in Schwarzschild ringdown}",
    eprint = "2606.07173",
    archivePrefix = "arXiv",
    primaryClass = "gr-qc",
    month = "6",
    year = "2026"
}

@article{Dyer:2026yex,
    author = "Dyer, Richard and Chung, Adrian Ka-Wai and Moore, Christopher J.",
    title = "{Modeling Direct Waves in Binary Black Hole Ringdowns}",
    eprint = "2606.25021",
    archivePrefix = "arXiv",
    primaryClass = "gr-qc",
    month = "6",
    year = "2026"
}

@article{Kankani:2026kst,
    author = "Kankani, Anuj and McWilliams, Sean T.",
    title = "{The Direct Wave is Not a Meaningful Test of Horizon Properties}",
    eprint = "2607.02380",
    archivePrefix = "arXiv",
    primaryClass = "gr-qc",
    month = "7",
    year = "2026"
}

@article{Sun:2026mto,
    author = "Sun, Ling and Ma, Sizheng and Chen, Patrick P. and Laeuger, Andrew and Lu, Neil and Chen, Yanbei",
    title = "{Complex frequency evolution of direct waves from binary black hole mergers}",
    eprint = "2608.23209",
    archivePrefix = "arXiv",
    primaryClass = "gr-qc",
    month = "8",
    year = "2026"
}

@article{Dyer:2026sox,
    author = "Dyer, Richard and Kankani, Anuj and Marks, Gareth Arturo and McWilliams, Sean T. and Moore, Christopher J.",
    title = "{Moving Beyond Quasinormal Modes and Reinterpreting Direct Waves in Binary Black Hole Merger-Ringdowns}",
    eprint = "2610.25176",
    archivePrefix = "arXiv",
    primaryClass = "gr-qc",
    month = "10",
    year = "2026"
}

@article{Kubota:2026qjh,
    author = "Kubota, Kei-ichiro and Manita, Yusuke and Motohashi, Hayato and Watarai, Daiki",
    title = "{Exact Cancellation of Horizon Modes in Kerr Ringdown}",
    eprint = "2609.07326",
    archivePrefix = "arXiv",
    primaryClass = "gr-qc",
    reportNumber = "RESCEU-23/26",
    month = "9",
    year = "2026"
}

@article{Bhagwat:2017tkm,
    author = "Bhagwat, Swetha and Okounkova, Maria and Ballmer, Stefan W. and Brown, Duncan A. and Giesler, Matthew and Scheel, Mark A. and Teukolsky, Saul A.",
    title = "{On choosing the start time of binary black hole ringdowns}",
    eprint = "1711.00926",
    archivePrefix = "arXiv",
    primaryClass = "gr-qc",
    doi = "10.1103/PhysRevD.97.104065",
    journal = "Phys. Rev. D",
    volume = "97",
    number = "10",
    pages = "104065",
    year = "2018"
}

@article{Kuntz:2026crc,
    author = "Kuntz, Adrien",
    title = "{The Start Time of Ringdown}",
    eprint = "2611.05472",
    archivePrefix = "arXiv",
    primaryClass = "gr-qc",
    month = "11",
    year = "2026"
}

@article{Nollert:1998ys,
    author = "Nollert, Hans-Peter and Price, Richard H.",
    title = "{Quantifying excitations of quasinormal mode systems}",
    eprint = "gr-qc/9810074",
    archivePrefix = "arXiv",
    doi = "10.1063/1.532698",
    journal = "J. Math. Phys.",
    volume = "40",
    pages = "980--1010",
    year = "1999"
}

@article{Glampedakis:2001js,
    author = "Glampedakis, Kostas and Andersson, Nils",
    title = "{Late time dynamics of rapidly rotating black holes}",
    eprint = "gr-qc/0103054",
    archivePrefix = "arXiv",
    doi = "10.1103/PhysRevD.64.104021",
    journal = "Phys. Rev. D",
    volume = "64",
    pages = "104021",
    year = "2001"
}

@article{Motohashi:2026mbn,
    author = "Motohashi, Hayato and Suichi, Yuto",
    title = "{Pole structure of the Kerr Green{\textquoteright}s function}",
    eprint = "2605.01964",
    archivePrefix = "arXiv",
    primaryClass = "gr-qc",
    doi = "10.1103/jlrh-3jpr",
    journal = "Phys. Rev. D",
    volume = "114",
    number = "4",
    pages = "044022",
    year = "2026"
}

@article{Arnaudo:2025uos,
    author = "Arnaudo, Paolo and Carballo, Javier and Withers, Benjamin",
    title = "{Beyond quasinormal modes: A complete mode decomposition of black hole perturbations}",
    eprint = "2510.18956",
    archivePrefix = "arXiv",
    primaryClass = "gr-qc",
    doi = "10.1103/1bcl-q79x",
    journal = "Phys. Rev. D",
    volume = "114",
    number = "2",
    pages = "024025",
    year = "2026"
}

@article{Su:2026fvj,
    author = "Su, Junquan and Khera, Neev and Casals, Marc and Ma, Sizheng and Chowdhuri, Abhishek and Yang, Huan",
    title = "{Decomposition of the Schwarzschild Green{\textquoteright}s function}",
    eprint = "2601.22015",
    archivePrefix = "arXiv",
    primaryClass = "gr-qc",
    doi = "10.1103/3n8r-8kbb",
    journal = "Phys. Rev. D",
    volume = "113",
    number = "10",
    pages = "104013",
    year = "2026"
}

@article{Rosato:2026moe,
    author = "Rosato, Romeo Felice and De Amicis, Marina and Pani, Paolo",
    title = "{Singular structures and causality of the Schwarzschild Green{\textquoteright}s function in the frequency domain}",
    eprint = "2603.20490",
    archivePrefix = "arXiv",
    primaryClass = "gr-qc",
    doi = "10.1103/58mx-v2wz",
    journal = "Phys. Rev. D",
    volume = "114",
    number = "2",
    pages = "024024",
    year = "2026"
}

\end{document}